\documentclass[conference]{IEEEtran}
\input epsf
\usepackage{graphicx}
\usepackage{caption}
\usepackage{floatrow}
\usepackage{amssymb}
\usepackage{todonotes}
\usepackage{array}
\usepackage{blindtext}
\usepackage{enumitem}
\usepackage{hyperref}
\usepackage{cite}
\usepackage{amsmath,amssymb,amsfonts}
\usepackage{algorithmic}
\usepackage{graphicx}
\usepackage{textcomp}
\usepackage{balance}
\usepackage{makecell}
\usepackage{url}
\usepackage{multirow}
\usepackage{xcolor}  
\begin{document}

\title{ A Survey  of Trust Management  \\ for Internet of Things} 
\author{\authorblockN{Alyzia Maria Konsta}
\authorblockA{Department of Applied Mathematics \\ and Computer Science \\
Technical University of Denmark\\
Email: akon@dtu.dk}
\and
\authorblockN{Alberto Lluch Lafuente }
\authorblockA{Department of Applied Mathematics \\ and Computer Science \\
Technical University of Denmark\\
Email: albl@dtu.dk}
\and
\authorblockN{Nicola Dragoni}
\authorblockA{Department of Applied Mathematics \\ and Computer Science \\
Technical University of Denmark\\
Email: ndra@dtu.dk}
}




%


\maketitle

\begin{abstract}
Internet of Things (IoT) is a network of devices that communicate with each other through the internet and provides intelligence to industry and people. These devices are running in potentially hostile environments, so the need for security is critical. Trust Management aims to ensure the reliability of the network by assigning a trust value in every node indicating its trust level. In this paper we systematically review and analyse the current state of Trust Management for IoT, we present a classification based on the methods used in every work and we discuss the open challenges and future research directions. 
\end{abstract}
\IEEEoverridecommandlockouts

%
\IEEEpeerreviewmaketitle

\section{Introduction}
\label{sec:introduction}
\PARstart{I}{nternet} of things (IoT) is a recent technology broadly used in our everyday lives. According to Cisco's annual report (2018-2023), the number of devices connected to IP networks will be more than three times the global population by 2023~\cite{Singh}. The term IoT was first introduced in 1999 by Kevin Ashton in the context of supply chain management ~\cite{GUBBI20131645}, but in the last decade, the concept has been used in multiple fields like agriculture~\cite{Bansal2020}, health care~\cite{Lin2021}, energy~\cite{Bansalen} and transportation~\cite{Bansaltr} among others. IoT forms a network of devices -such as RFID, sensors, mobile phones, etc.- that are communicating through the internet. These devices gather information from their environment and provide intelligence to industry and people. 

IoT objects are running in remote locations in potentially hostile environments, so they are vulnerable to security attacks. However, these resource-constrained devices cannot support the customary security algorithms, which require powerful hardware and software. Taking into account the magnitude of IoT and the domains using this technology, we can imagine that a huge amount of sensitive information is processed by IoT devices. Therefore, the need for security is crucial. One method used to assess the reliability of the network is trust management. Trust management aims to ensure the reliability of the network by assigning a trust value to every node, indicating its trust level. Thus, the information provided by a node with a high trust level is considered reliable. To create a trust relationship, at least two entities must be involved: the trustor and the trustee.

In this work, we present an exhaustive survey of trust management for IoT. This study provides a classification based on the methods and technologies used for trust formation. In every paper, we examine nine different dimensions, including limitations and strengths.

\paragraph*{Scope of the Paper} 
The scope of the paper focuses on specific aspects of trust management. In particular, how information necessary for trust is gathered, how trust is updated, how trust is formed, how trust information is propagated among nodes, which is the threat model considered, whether the approach is validated with experiments and how those are conducted, and whether tools like simulators are available to support evaluation of approaches to trust management.  Further information on these aspects is provided in Section~\ref{class}.  We decided to include these aspects of trust after examining which of them are included in the current literature, to which we added the experiment and simulator aspects since we believe it is important to consider tool-based support to assess the proposed approaches. In our perspective, it is important to examine \textit{which} experiments were conducted and \textit{how} they were conducted in every work. We also discuss the limitations and strengths of each paper.

\paragraph*{Goal of the Paper}
The goal of this survey is to identify existing trust management mechanisms and to classify them based on their main technologies in order to spot limitations in each category. We highlight limitations and research gaps by examining some key aspects of trust. Our objective is to provide an overview of the field and discuss the challenges identified. We hope that this will help interested readers navigate the vast literature in this field.  We also aim to help the reader design a solid trust management system and navigate through the literature based on the method used and some key properties.



\paragraph*{Contributions}
The main contributions of this paper are:
\begin{enumerate}
\item A literature review of the existing trust management techniques.
\item A categorization of the existing literature based on the techniques/tools used to form the trust management method.
\item Pointing out current challenges and future research directions for trust management in the IoT.
\end{enumerate}

\paragraph*{Structure of the Paper}
The rest of the paper is structured as follows:
\begin{itemize}
\item{Section~\ref{back}:} Provides the necessary background by summarizing the main concepts of IoT and trust management used throughout the paper. 
\item{Section~\ref{research}:} Presents the research method we have followed to structure our research and the main questions we aim to address through this survey. 
 \item{Section~\ref{related}:} Discusses the related works, i.e. other surveys that have considered the topic of trust management in IoT. 
\item{Section~\ref{overview}:} Presents an overview of the categories by providing the reader with a quantification study. 
\item{Section~\ref{class}:} Provides the classification of the literature and description for all the papers participating in this research. 
\item{Section~\ref{challenges}:} Discusses the main challenges and future research directions.
\item{Section~\ref{conclusion}:} Summarizes and concludes the study.
\end{itemize}



\section{Background}\label{back}
In this section, we provide an overview of the basic concept we are going to discuss in the following chapters.
The structure of this section is as follows:
\begin{itemize}
\item Section~\ref{sec:intro-IoT} provides an introduction to the IoT. 
\item Section~\ref{sec:intro-trust} provides an overview of the main concepts of trust management. 
\item Section~\ref{sec:intro-arch} recalls the classical three-layer architecture of the IoT. 
\item Section~\ref{sec:intro-cat} describes the main categories used to structure our survey. 
\item Section~\ref{sec:intro-attacks} provides an overview of the main classes of attacks on IoT as known from the literature. 
\item Section~\ref{sec:intro-cloud}, Section~\ref{sec:intro-edge} and Section~\ref{sec:intro-block} respectively cover three technologies that are relevant for many of the papers included in the survey, namely cloud computing, edge computing, and blockchain. 
\end{itemize}

\subsection{Introduction to IoT} \label{sec:intro-IoT}

Internet of Things (IoT) refers to a network of devices that can communicate with each other, and collect and exchange data over the internet. In effect, IoT enables everyday objects to become ``smart'' enhancing their functionality and enabling new services. IoT surrounds a diverse ecosystem of devices ranging from everyday items like thermostats, and refrigerators to complex machinery, equipped with sensors for data collection.

This recent technology has a significant impact on our everyday lives and industries. The data collected by IoT devices can be analyzed to derive insights, make predictions, and optimize processes. Machine learning and AI can play a role in extracting valuable information from IoT data. IoT has a wide range of applications across various industries, including smart homes, healthcare, agriculture, manufacturing, transportation, energy management, and more. Examples include smart cities, wearable fitness trackers, and autonomous vehicles.

IoT has the potential to revolutionize many aspects of our lives and industries by providing real-time data, automation, and insights that can lead to improved efficiency, convenience, safety, and sustainability. However, it also presents challenges, such as data security. Since the number of connected devices grows drastically and IoT devices are operating in potentially hostile environments the need for security is crucial. Moreover, IoT devices are usually resource-constrained devices that cannot support the broadly used security algorithms that consume a vast amount of energy and resources. Thus, the need for alternative security mechanisms that support the characteristics of these devices arises. One mechanism that is broadly used is trust management\cite{YAN2014120, Singh}.

\subsection{Trust Management}  \label{sec:intro-trust}

 In a dynamic environment such as an IoT network, it is really important to detect malicious nodes. In our everyday lives, we want to collaborate and connect with trustworthy people, both in our personal and professional lives. In the same way, IoT devices, in order to function smoothly, need to interact with trustworthy nodes, that provide them with honest information.

The basis of trust is the accurate identification of IoT devices. Each device should have a unique identity so that it can be distinguished in the network. With this unique identity, the device would be able to access the network and communicate with other devices. After the authentication is in place, access control mechanisms define which actions and data each device is authorized to access. 

It is also crucial to ensure data integrity and confidentiality, both in transit and at rest, from unauthorized access and tampering. Trustworthy devices are essential for the smooth functionality and security of IoT networks. This includes assessing the behavior of the devices, monitoring suspicious anomalies, and revoking trust if necessary. The process starts with collecting information for other nodes, using this information to compute the trust level, and then storing and using this piece of information. The trust level determines if a node can trust or interact with another. As people, we would never close a deal with a person that we did not trust; the same applies to the IoT nodes in a network. Different trust models can be used in IoT, such as hierarchical trust models, reputation-based trust models, and trust models, depending on the specific IoT application and architecture. Trust management in IoT is an ongoing process that requires collaboration between manufacturers, service providers, and end users to establish and maintain trust throughout the life of IoT devices and services.

Trust management is associated with multiple challenges, such as the cold-start problem, which is the initialization of the trust values during bootstrapping or a new node entering a network. Sometimes privacy issues may arise since trust management mechanisms may handle sensitive data and the resources that have to be consumed for them to operate\cite{Singh, AlshehriSUR, SHARMA2020475}. In Section~\ref{challenges} we are analyzing and summarizing all the challenges identified throughout this survey.

\subsection{IoT Architecture}  \label{sec:intro-arch}
According to most researchers, IoT is a three-layer architecture \cite{Jab} \cite{KUMAR2021}. These are the Perception layer, the Network layer, and the Application layer.
\begin{itemize}
    \item \textit{Perception layer:} It consists of physical devices, such as sensors, that gather information from the environment.
    \item \textit{Network layer:} This layer is responsible for connecting the devices to servers and network devices. Also, the protocols of this layer are used to transmit and exchange information among the devices.
    \item \textit{Application layer:} Serves as an intermediate layer between the network and the IoT services. The data collected from the smart devices is transferred to the application layer. Applications like smart health, smart home, etc. belong to this layer.
\end{itemize}


\subsection{Categories}  \label{sec:intro-cat}
One of the contributions of this work is to point out the tools, methods, and technologies being used to form trust management techniques. The categories defined represent the technologies used in each paper. We divide the papers into nine different categories: Blockchain, Context, Social, Game Theory, Probabilistic, Prediction, Fuzzy, Direct, and Recommendations. These categories represent the technologies used in every paper to gather the information for the trust formation, compute the final trust, and store the trust-related data as stated in Table \ref{tab:categories}. The rows in Table \ref{tab:categories} represent the categories defined, and the columns the aspects of trust management to every category contributes.  

\begin{table}[!t]\scriptsize
 \caption{Categories used in this work}
 \label{tab:categories}
\resizebox{\textwidth}{!}{ \begin{tabular}{||c c c c||} 
 \hline
 Category & Info Gathering   & Computation & Storing \\ [0.5ex] 
 \hline\hline
 Blockchain &  & \checkmark & \checkmark \\ 
 Context &  & \checkmark &  \\
 Social &  & \checkmark &  \\
 Game Theory &  & \checkmark &  \\
 Probabilistic &  & \checkmark &  \\
 Prediction &  & \checkmark &  \\
 Fuzzy &  & \checkmark &  \\
 Direct & \checkmark &  &  \\
 Recommendations & \checkmark &  &  \\ [1ex] 
 \hline
 \end{tabular}}
\end{table}

\begin{figure}[!t]
\centering
\includegraphics[width=\textwidth]{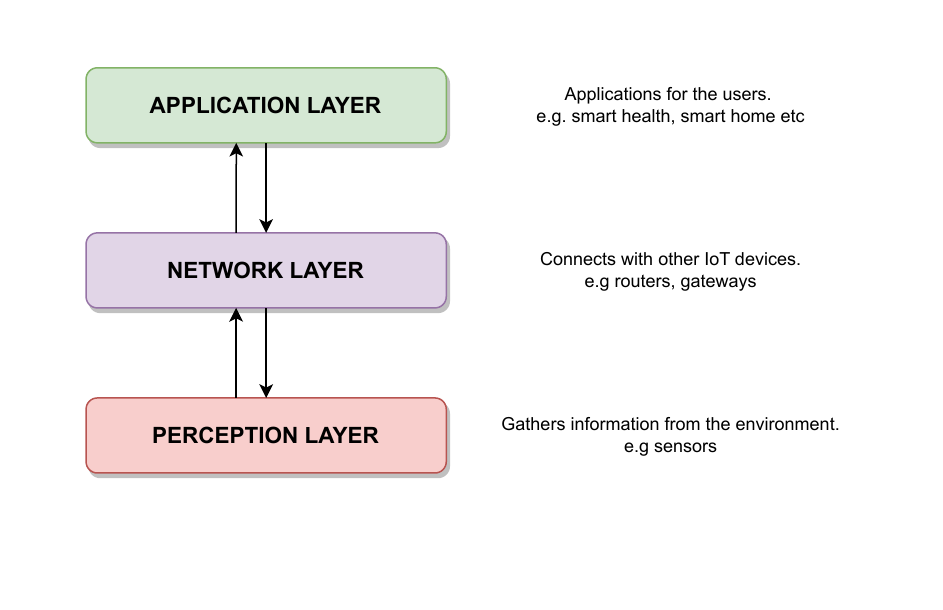}
\caption{IoT Architecture}
\label{arch}
\end{figure}

To form a trust management technique one can combine the tools offered by any category; thus, each paper could in principle be part of more than one category.

We are going to examine seven different dimensions in every category to identify the research gaps: information gathering, experiments, trust propagation, threat model, trust update, trust formation, and the simulator. We summarize some limitations and strengths for each dimension in Table~\ref{tab:lim}. Following we introduce the seven dimensions of trust: 

\subsubsection{Information Gathering}
One trust management system should collect data to execute the trust computation.
Information Gathering is the first step toward trust computation. It refers to the process of collecting knowledge about the trust parameters. We can divide information gathering into two categories:
\begin{itemize}
    \item \textit{Direct Trust:} Trust is formed based on direct observations and interactions between the two parties involved in the trust relationship.
    
    \item \textit{Indirect Trust:} The trustor and the trustee do not share any previous interactions. Trust is formed based on the recommendations of other nodes \cite{SHARMA2020475, GUO20171}. 
\end{itemize}

\subsubsection{Experiments}
We are also going to examine what kind of experiments took place in every work.

\subsubsection{Trust Propagation} Trust propagation refers to how the trust evidence is propagated to the nodes involved in the system, and it is divided into two categories:

\begin{itemize}
   \item \textit{Distributed:} Every node stores the trust values of the other nodes. The nodes interact and exchange trust evidence with other nodes. The nodes independently store and compute the trust values without involving a central authority in the procedure. 
   
   \item \textit{Centralized:} A central authority is present to compute and store the trust values. This entity is responsible for propagating and handling the trust evidence \cite{SHARMA2020475, GUO20171}.
\end{itemize}

\subsubsection{Threat Model:} It is a set of attacks examined in each work. The malicious node can perform these attacks in the system under investigation.

\subsubsection{Trust update} Trust update refers to when the trust is updated and can be divided into two categories:

\begin{itemize}
    \item \textit{Event driven:} When a specific event triggers the trust update. For example, when a node requests an interaction, trust update can be triggered.
    \item \textit{Time driven:} When the trust is being updated in time intervals \cite{SHARMA2020475, GUO20171}. 
\end{itemize}

\subsubsection{Trust Formation}
 Trust formation refers to how the overall trust is formed. It is divided into two categories:
 
 \begin{itemize}
     \item \textit{Single Trust:} Only one trust parameter is considered to form the trust of a node.
     \item \textit{Multi Trust:} Several parameters are used to form the trust of a node since the trust is considered multidimensional \cite{SHARMA2020475, GUO20171}.
 \end{itemize}
 
\subsubsection{Simulator:} Refers to the simulator used in every work to perform the experiments. Of course, in some cases, real devices were used. 


\begin{table*}[ht] \scriptsize
\caption{Limitations and Strengths for each dimension}
  \label{tab:lim}
 \resizebox{\textwidth}{!}{   \begin{tabular}{| m{2cm} | m{8cm} | m{8cm} |}
        \hline
        Dimension & Limitations  & Strengths\\
        \hline
        Direct Trust & 
        \begin{itemize} 
            \item A malicious node might act malevolent to other nodes.
        \end{itemize} &
        \begin{itemize}
            \item Based on your own experience.
        \end{itemize}  \\
        \hline
        Indirect Trust &
        \begin{itemize} 
            \item Some nodes might provide false recommendations.
            \item Need for filtering recommendations.
        \end{itemize} &
        \begin{itemize}
            \item Provides a global view of the nodes' behaviors.
        \end{itemize} \\
        \hline
        Experiments &
        \begin{itemize} 
            \item No limitation on adding experiments.
        \end{itemize} &
        \begin{itemize} 
            \item Experiments can support the methodology and its functionality.
            \item Key metrics prove how the method improves existing approaches.
        \end{itemize} \\
        \hline
        Distributed &
        \begin{itemize} 
            \item More resources are consumed, for every node to store and compute the trust values.
        \end{itemize} &
        \begin{itemize} 
            \item Every node holds its values and no central authority can provide false data.
        \end{itemize} \\
        \hline
        Centralized &
        \begin{itemize} 
            \item A central authority can provide false data. 
            \item Static procedure.
        \end{itemize} &
        \begin{itemize} 
            \item The computational and storage weight is taken away from the nodes.
        \end{itemize} \\
        \hline
        Threat Model &
        \begin{itemize} 
            \item No limitation on stating the threat model.
        \end{itemize} &
        \begin{itemize} 
            \item Understand where every method is applicable.
            \item A key metric to compare approaches.
        \end{itemize} \\
        \hline
        Event-Driven &
        \begin{itemize} 
            \item Very often updates can lead to more energy consumption.
        \end{itemize} &
         \begin{itemize} 
            \item The trust is updated after every interaction and a malicious node will be caught immediately.
        \end{itemize} \\
        \hline
        Time-Driven &
        \begin{itemize} 
            \item For long intervals, a malicious node can act for a certain time.
        \end{itemize} &
         \begin{itemize} 
            \item A nice balance between frequency and saving resources can save a lot of energy.
        \end{itemize} \\
        \hline
        Single-Trust &
        \begin{itemize} 
            \item One view of trust.
        \end{itemize} &
         \begin{itemize} 
            \item Saving energy.
        \end{itemize} \\
        \hline
        Multi-Trust &
        \begin{itemize} 
            \item More energy is needed to compute all parameters.
        \end{itemize} &
         \begin{itemize} 
            \item Multidimensional view of trust.
        \end{itemize} \\
        \hline
        Simulator &
        \begin{itemize} 
            \item No limitation on stating the setup environment.
        \end{itemize} &
         \begin{itemize} 
            \item Details about the setup of an experiment can hold a lot of information (e.g. simulation vs. real data).
        \end{itemize} \\
        \hline
    \end{tabular}}
\end{table*}

 \subsection{Attacks}  \label{sec:intro-attacks}
Based on the examined literature, an IoT node can perform the attacks presented in Table \ref{tab:acategories}. An attack can be trust-related or belong to a different layer of the IoT architecture.

 In an IoT system using a trust management technique where the nodes are evaluated, the trust level of the node plays an important role in their image. A node with a good trust level can have multiple collaborators and influence in the system. Hence, we are concerned with trust-related attacks. First, we are going to elaborate on these kinds of attacks, identified so far in the literature \cite{GUO20171,Saeed}: 
 
 \begin{table}[h!]\scriptsize
 \caption{Category of every attack}
 \label{tab:acategories}
 \resizebox{\textwidth}{!}{\begin{tabular}{||c c c c c||} 
 \hline
 Attack & Trust   & Application & Network & Perception \\ [0.5ex] 
 \hline\hline
 BMA & \checkmark &  &  & \\ 
 BSA &  \checkmark &  &  & \\
 SPA &  \checkmark &  &  & \\  
 OSA &  \checkmark &  &  & \\ 
 OOA &  \checkmark &  &  & \\ 
 EA &  &  &  & \\ 
 NCA &  & \checkmark &  & \\ 
 RA &  &  &  & \checkmark \\
 SA &  &  & \checkmark &  \\
 WA &  &  &  &  \\
 DoS &  &  & \checkmark & \\
 SFA &  &  & \checkmark & \\
 BA &  &  & \checkmark & \\
 WHA &  &  & \checkmark & \\
 IA &  & \checkmark &  &  \\
 SDA & &  & &  \\ [1ex] 
 \hline
 \end{tabular}}
\end{table}

 \begin{itemize}
     \item \textit{Bad mouthing attacks (BMA):} A malicious node can provide bad recommendations for an honest node, trying to ruin its reputation. The goal of this attack is to lower the reputation of an honest node.
     \item \textit{Ballot stuffing attacks (BSA):} A malicious node is providing good recommendations for other malicious nodes. The goal of this attack is to increase the trust level of other malicious nodes, thus increasing their influence in the network.
     \item \textit{Self-promoting attacks (SPA):} A malicious node can provide good recommendations for itself to increase its influence in the network.
     \item \textit{Opportunistic service attacks (OSA):} A malicious node can provide good service to gain a high trust level and then cooperate with other malicious nodes to perform bad-mouthing and ballot-stuffing attacks.
     \item \textit{On-Off attacks (OOA):} A malicious node can provide sometimes bad and sometimes good services. With this attack, a node avoids being labeled as an untrustworthy node. 
 \end{itemize}
 
 In the scope of this paper, we are also going to discuss the following types of attacks mentioned in the literature included in this survey:
 
 \begin{itemize}
     \item \textit{Eclipse attack (EA):} In this type of attack, the malicious node isolates the victim from the rest of the network \cite{DEDE}.
     
     \item \textit{Node Capture attack (NCA):} This type of attack targets the physical devices of the IoT, in terms of communication links, fake data input, etc. \cite{KUMAR2021}
     
     \item \textit{Replay attack (RA):} In this type of attack, the malicious node is listening to a communication to gain information and misdirect the receiver. \cite{KUMAR2021} For example, if Alice shares a piece of information with Bob (to prove her identity), then Eve is eavesdropping on the conversation and stores the information Alice shared. Now Eve can maliciously communicate with Bob and pretend to be Alice. 
     
     \item \textit{Sybil attack (SA):} A malicious node has multiple identities and can place itself simultaneously in different places in the network.
     
     \item \textit{Whitewashing attack (WA):} When a node with a bad reputation is re-entering the network with a different identity to reset its reputation.
     
     \item \textit{Denial of Service attack (DoS):} When a malicious node is sending multiple requests to the network to make it unavailable for the rest of the users.
     
     \item \textit{Spoofing attack (SFA):} When a node uses a different identity and pretends to be someone else.
     
     \item \textit{Blackhole attack (BA):} When a node is deleting all messages, it is supposed to forward. This attack is creating a gap in the network.
     
     \item \textit{Wormhole attack (WHA):} The nodes involved in this attack are stronger nodes that communicate at longer distances. The packets are forwarded from one malicious node to the other through a tunnel. In this way, they can trick the other nodes of the network into believing that these two nodes are closer.
     
     \item \textit{Injection attacks (IA):} A malicious code is injected to disturb the smooth functioning of the network.
     
     \item \textit{Sleep deprivation attack (SDA):} The malicious node is making frequent requests to a node to keep it awake and consume all the battery resources quickly \cite{Pirretti}.
     
 \end{itemize}

 \subsection{Cloud Computing}  \label{sec:intro-cloud}

 Cloud computing enables users and devices to store and process huge amounts of data with the use of services provided on the internet. Some of the benefits of cloud computing:

 \begin{itemize}
     \item Cost: No need to acquire specialized hardware or IT teams to manage these infrastructures.
     \item Performance: Provide high-performance services to heterogeneous devices.
     \item Speed: High-speed services.
     \item Security: Provides services that enhance security.
 \end{itemize}

For the reasons mentioned above, cloud computing can be paired with IoT devices. IoT devices collect a vast amount of information and send it to the cloud for storage and computation of different metrics. The cloud is on the network layer Figure~\ref{arch}. The data are collected by the perception layer and sent to the network layer, where the cloud services process the data and reform them to be useful for the users on the application layer~\cite{cloud}.

\subsection{Edge Computing}  \label{sec:intro-edge}

Cloud computing, discussed above, offers some major advancements, but it remains far away from the local network. This may result in delays, poor raw data, and no real-time insights. These drawbacks led the research community to come up with Edge Computing. This framework allows the IoT gateways to perform pre-processing on the raw data and send only the useful ones to the cloud. So, edge computing can be seen as a way to store and process data closer to data production, decreasing delays~\cite{edge}.

\begin{figure}[!t]
\centering
\includegraphics[width=3.5in]{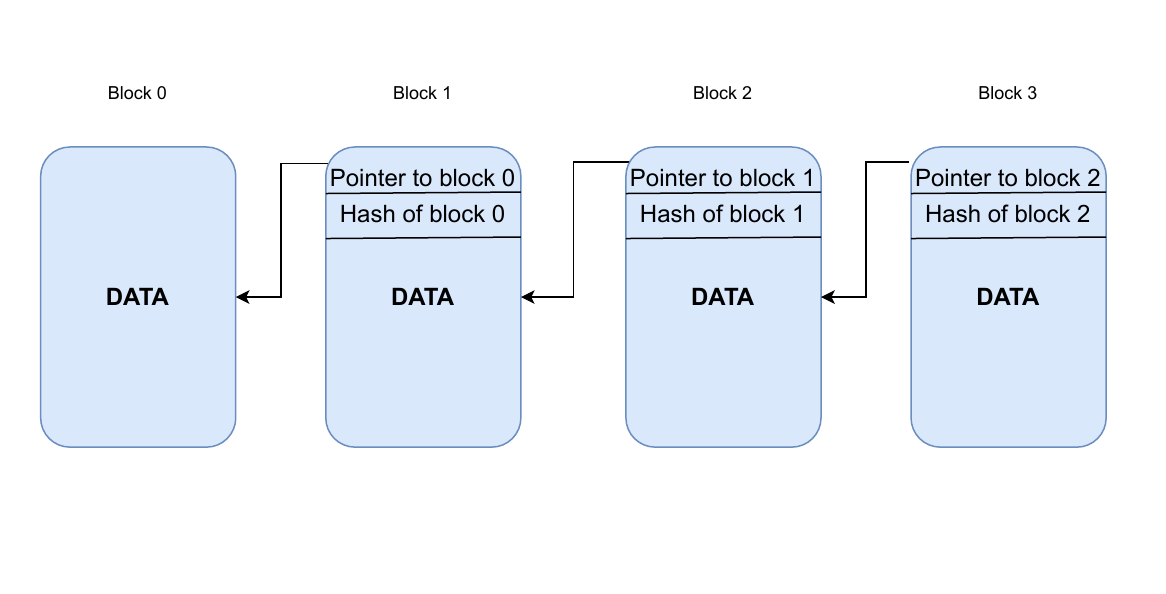}
\caption{Blockchain Example}
\label{blockchain}
\end{figure}

\subsection{Blockchain}  \label{sec:intro-block}

The blockchain is a decentralized, distributed ledger consisting of a list of records called blocks. These blocks are linked using a cryptographic hash, as seen in Figure~\ref{blockchain}. The first application of blockchain was for the Bitcoin cryptocurrency, but it has also been used in many other applications. 

The most significant properties of blockchain are transparency, immutability, and decentralization.

Transparency means that all users can see other transactions only through their public addresses. Immutability means that once a piece of information is stored on the blockchain, it cannot be changed. Decentralization means that everything is distributed and there is no central authority~\cite{block}. 

\section{Research Method}\label{research}
In this section, we present the research method we adopted to discover the existing literature on trust management in IoT. We followed the research method proposed by Petersen et al \cite{PETERSEN20151}. In the following, we describe the research questions, the search method, and the study selection. The structure of this section is as follows:
\begin{itemize}
\item Section~\ref{sec:researchquest} defines the research questions.
\item Section~\ref{sec:searchmethod} provides the search method we followed.
\item Section~\ref{sec:studyselection} provides the exclusion criteria we followed during study selection. 
\end{itemize}

\subsection{Research Questions}\label{sec:researchquest}
This paper aims to study the existing literature on trust management in IoT, so we focus on the following research questions:

\begin{itemize}
    \item \textbf{RQ1:} Which methods are currently used in the field?
    \item \textbf{RQ2:} What is the threat model of those proposals?
    \item \textbf{RQ3:} What are the strengths and limitations of each proposal?
    \item \textbf{RQ4:} What are the challenges and future directions?
\end{itemize}

\subsection{Search Method}\label{sec:searchmethod}
We used the PICOC criteria \cite{kitchenham2007guidelines} to come up with relevant keywords for our search:

\begin{itemize}
    \item \textbf{Population:} We are interested in works in IoT nodes.
    \item \textbf{Intervention:} We are interested in works that propose a trust management technique.
    \item \textbf{Comparison:} We compare different kinds of trust management schemes in the IoT based on design features, security capabilities, and performance.
    \item \textbf{Outcomes:} We present trust management techniques for the IoT: opportunities and limitations, as well as future challenges.
    \item \textbf{Context:} We are interested in any paper that proposes a trust management technique for IoT nodes.
\end{itemize}

Based on the above criteria, we came up with the following keywords: ``Internet of Things'', ``IoT'', ``trust'', ``trustworthy'' and ``node''. We performed our search in DTU Findit \url{https://findit.dtu.dk}, which is an open (guest access) database that includes publications from widely known journals and databases: Elsevier, IEEExplore, ACM Digital Library, etc. We used the following query: title:(IoT OR "internet of things") AND title:(trust OR trustworthy) AND abstract:(node). Our search returned 341 papers. The final pool of papers was selected based on the study selection we describe in the next paragraph.

\subsection{Study Selection}\label{sec:studyselection}
We started with 341 papers after identifying the duplicates, we applied the following exclusion criteria: \begin{itemize}
    \item \textbf{E1:} The full text of the paper is not available. 
    \item \textbf{E2:} The papers in not provided in English.
    \item \textbf{E3:} The work is not focusing on IoT.
    \item \textbf{E4:} The work is not focused on trust management techniques.
\end{itemize}

After applying the above exclusion criteria, we also performed the snowballing \cite{snow} technique and ended up with 53 papers, which constitute the final pool.

\section{Related Work}\label{related}
In this section, we present some papers conducting surveys on trust management for the IoT. We also present a table with useful information regarding the papers. On Table~\ref{tab:RW} we included 8 columns. The first column indicates the paper under examination. The rest of the columns refer to different kinds of characteristics of each paper. When a characteristic denoted by the corresponding column is satisfied, we fill out a $\checkmark$. The second column, Trust Attacks, describes if a paper takes into account trust-related attacks and the third column describes if the paper takes into account other types of attacks. The fourth column denotes if a paper is classifying the papers under study based on the underlying technology used in each one to produce the trust results. The fifth column indicates if a paper is published after 2020. The sixth column General indicates if the paper is conducting a general survey on trust management or focusing on a specific method or field. The seventh column, Future Directions denotes if a paper is proposing future directions for the research community. Finally, the last column indicates if the paper proposes some solution for the challenges identified.

Guo et al. \cite{GUO20171} classified trust computation models for service management in IoT systems. Their classification is based on the techniques used in trust composition. They proposed five design dimensions for a
trust computation model: trust composition, trust propagation, trust aggregation, trust update, and trust
formation. The authors mentioned the pros and cons of each dimension's options. Finally, they also identified gaps in IoT trust computation research and
suggested future research directions. 
This work was published before 2020, and the authors only mention trust-related attacks in their survey.

Yan et al. \cite{YAN2014120}  investigated the properties of trust, proposed objectives for IoT trust management, and
provided a survey of the literature on trustworthy IoT. Furthermore, the authors discussed
unsolved issues and research challenges, and proposed a research
model for holistic trust management in IoT. To conduct holistic IoT trust management, the trust properties that impact trust relationships were explored and
classified into five categories: Trustee's objective properties, Trustee's subjective properties, Trustor's subjective properties, Trustor's objective properties, and the Context in which the trust relationship resides.
This paper was published before 2020, and the authors do not mention the threat models of the papers included in their survey.

Singh et al. \cite{Singh} discussed the fog computing three-layer architecture and state-of-the-art models. The bottom layer of Fog Computing comprises IoT devices \cite{Singh}. Through their survey on trust management in Fog Computing Singh et al., also identified trust and security challenges.  
This recent work only focuses on Fog Computing architectures.

Kumar et al. \cite{KUMAR2021} focused on research regarding trust using Blockchain technology in the IoT environment. The authors pointed out some challenges and
issues of trust management in IoT and proposed Blockchain-based solutions. Furthermore, some issues regarding the integration of Blockchain with IoT were discussed. Finally, a comparative analysis between traditional and Blockchain-based trust management techniques was presented.
Kumar et al. only focus on blockchain-related research.

Sharma et al. \cite{SHARMA2020475} presented the different stages involved in the process
of trust management. Furthermore, the authors presented a survey on trust management schemes. The survey is conducted considering direct
observations and indirect recommendations, distributed, semi-distributed, centralized schemes, and blockchain-based schemes for trust management in IoT. Moreover, they provided
a comparative study of the existing schemes based on some system parameters like the computation model,
input attributes, evaluation tools, and performance metrics, examining their strengths and weaknesses.
The paper also highlights open research challenges and presents future directions
for the researchers.
Sharma et al. do not provide classification based on the technologies used and focus only on trust-related attacks.

Alshehri et al. \cite{AlshehriSUR} were focused on scalable and context-aware trust management for the IoT. They present the concept of IoT and the importance of trust. The authors also provided a comparative evaluation of existing trust solutions for the IoT focusing on scalability. Also, they presented a trust management protocol for the IoT. Furthermore, the authors also provided a context-aware evaluation of the IoT and compared the different trust solutions.
Finally, the authors gave some future directions for research.
This work was published before 2020, and the authors do not provide the threat models of the works included in this survey. Also, this work focuses only on context-aware trust management.

Saeed et al. \cite{Saeed} proposed a classification tree in this survey for trust management models. The classification scheme takes into account five dimensions of trust. They do not examine the experiments conducted in every work or the threat model mentioned in every category. The authors examine some trust-related attacks on IoT devices. Finally, they point out some future directions.
This work only mentions trust-related attacks; They do not classify the papers based on the technologies used.

Pourghebleh et al. \cite{Pourghebleh} presented a survey where the selected techniques were categorized into four
main classes, including recommendation-based, prediction-based,
policy-based, and reputation-based. The authors also present a discussion where they compare the literature based on some metrics, such as accuracy,
adaptability, availability, heterogeneity, integrity, privacy, reliability, and scalability. Furthermore, some future challenges and directions are provided.
Our work has more updated literature, covering papers published since 2020, while~\cite{Pourghebleh} covers only papers published until 2019. As a result, our work covers a total of 36 papers that were not covered before.
Moreover, while~\cite{Pourghebleh} 
focuses only on trust-related attacks, Our work considers additional classes of attacks at the application, network, and perception level. We argue that a well-designed trust management system should be capable of dealing with all kinds of attacks. Indeed, this position is shared by some of the surveyed papers, which consider threat models and concrete attack methods that are not directly trust-related. 

\begin{table*}[]\scriptsize
  \caption{Comparison with Related Work}
  \label{tab:RW}
  \centering
\resizebox{\textwidth}{!}{\begin{tabular}{||m{5em}| m{5em} | m{5em} | m{6em} | m{5em}| m{5em} | m{5em} | m{5em}||} 
 \hline
 Paper & Trust Attacks& Other Attacks & Methods' Classification  & After 2020 & General  & Future Directions & Solutions \\ [0.5ex] 
 \hline\hline
 \cite{GUO20171} & \checkmark &  &  & & \checkmark &\checkmark & \\ 
 \hline
 \cite{YAN2014120} &  &  &  & & \checkmark & \checkmark&\\
 \hline
 \cite{Singh} & \checkmark &  & \checkmark  & \checkmark& \checkmark& \checkmark & \\  
 \hline
  \cite{KUMAR2021} & \checkmark  & \checkmark  &   &  \checkmark & & \checkmark& \checkmark\\
 \hline
   \cite{SHARMA2020475} & \checkmark  &   &   &  \checkmark & \checkmark & \checkmark& \\
 \hline
 \cite{AlshehriSUR} &   &   &   &   &  &\checkmark & \\
 \hline
 \cite{Saeed} &  \checkmark &   &   &  \checkmark & \checkmark &\checkmark & \\
 \hline
  \cite{Pourghebleh} & \checkmark  &   &   &  & \checkmark& \checkmark & \\
 \hline
   Our work & \checkmark  & \checkmark  &  \checkmark & \checkmark & \checkmark& \checkmark & \checkmark \\[1ex] 
 \hline
 \end{tabular}}
\end{table*}

The summary of our findings is presented in Table \ref{tab:RW}. 
We can observe from the Table that most of the works refer only to trust-related attacks. To identify the weaknesses of the field, one should also take into account the non-trust-related attacks that are being tackled in the literature.

Until now, only \cite{Singh} refers to the technologies used to classify the papers but focuses only on trust-related attacks. Also, only \cite{KUMAR2021} provides some solutions to the challenges presented, but the survey is focused on Blockchain-based trust management techniques.

We can conclude that our work provides a more complete view of general trust management techniques. We point out the technologies used in every work and provide a categorization based on these technologies. We also take into account all kinds of attacks and provide future directions and a solution to support the design of a multidimensional trust management system. 

\section{OVERVIEW OF THE CATEGORIES}\label{overview}

In this section, we will provide a comprehensive picture of the categories and attacks mentioned. More specifically, we will conduct a quantification study and discussion on attacks, publishers, and publication years. The structure of this section is as follows:
\begin{itemize}
\item Section~\ref{sec:eachcategory} presents discusses the percentage of papers in each category.
\item Section~\ref{sec:precentageofattacks} presents data related to the trust model.
\item Section~\ref{sec:publisher} provides results regarding the publisher and the publication year.
\item Section~\ref{sec:attackstrustpro} provides more details about the categories.
\end{itemize}

\subsection{Percentage of papers in each category}\label{sec:eachcategory}
\begin{figure}[!t]
\includegraphics[width=\textwidth]{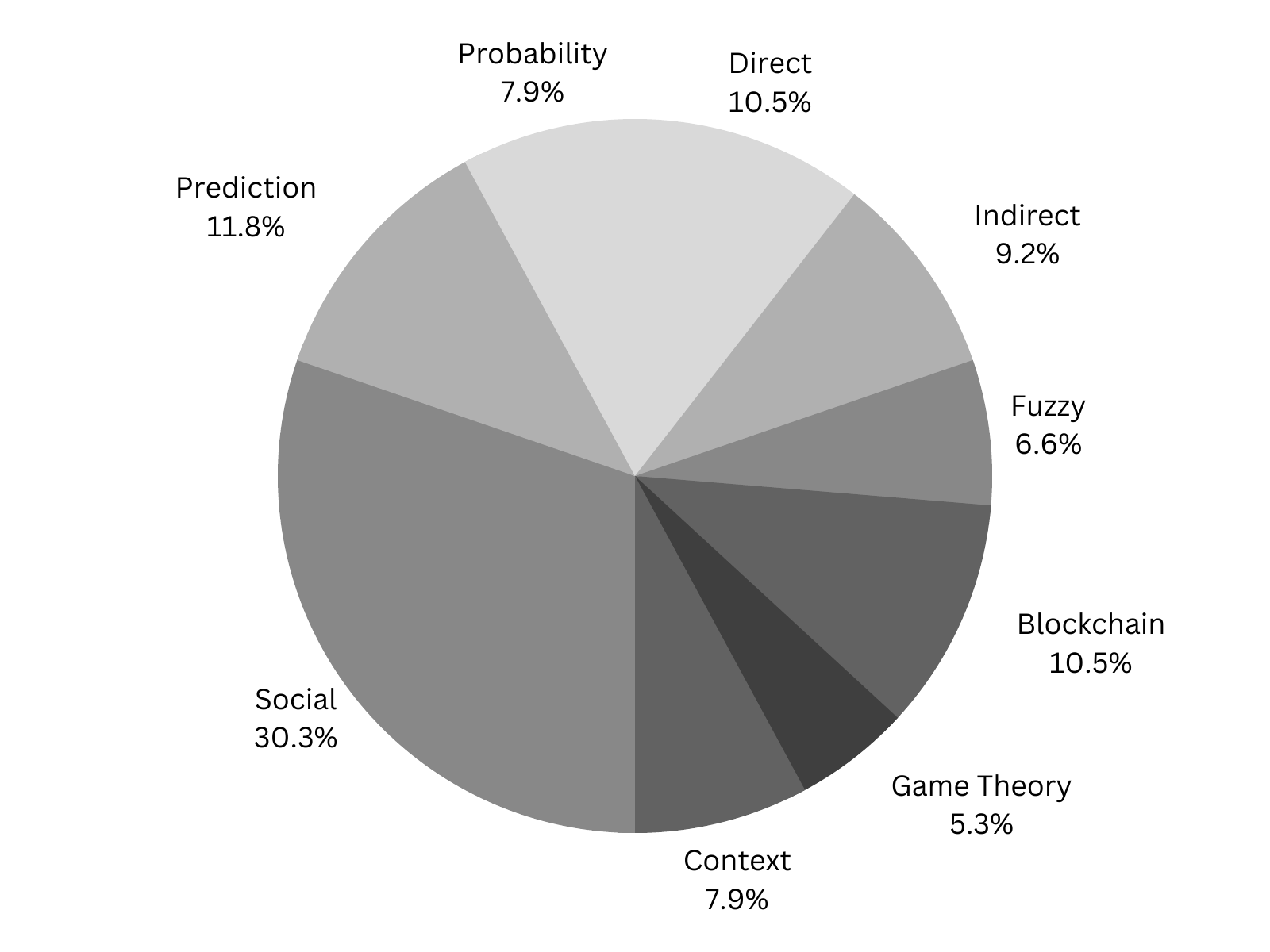}
\caption{Percentage of papers classified in each category}
\label{fig:class}
\end{figure}

In total, the survey consisted of 53 papers. Each paper may belong to different categories. We classified the papers into nine different categories. In Figure \ref{fig:class} you can see the percentage of papers classified in each category.

As we can see, the most popular category is Social. To calculate trust in this category, social aspects such as friendship are considered. Like in our everyday lives, we are more invested in trusting people that we have previously interacted with or have a social group in common. It simulates a network of nodes as a group of things with social interactions. The nodes can develop relationships with the other participants, and the level of trust is related to their social interactions. This scheme also provides the opportunity to filter the recommendations.

\subsection{Percentage of attacks}\label{sec:precentageofattacks}
\begin{figure}[!t]
\includegraphics[width=\textwidth]{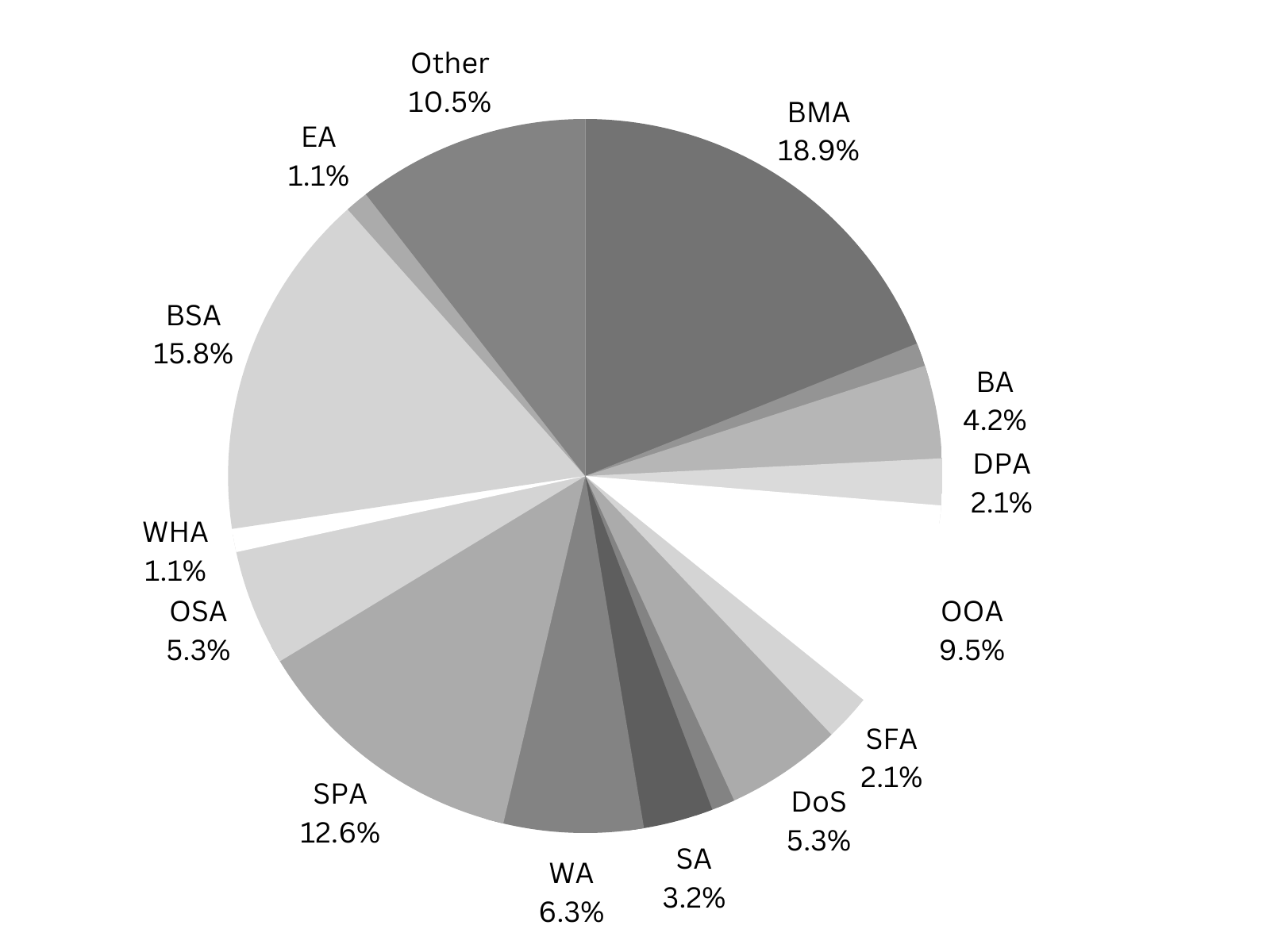}
\caption{Percentage of every attack mentioned in every threat model}
\label{fig:attacks}
\end{figure}

In this section, we present some data related to the threat model provided. It is important to mention that 28 out of the 53 papers defined a specific threat model for their research. In Figure \ref{fig:attacks} you can see the statistical information regarding the attacks studied in the literature.

We observe that trust-related attacks -BMA, BSA, SPA, OSA, OOA-  are studied the most in the literature. It is really important for a trust management system to be able to deal with trust-related attacks, but this is only the foundation. A well-designed trust management system should be capable of dealing with all kinds of attacks. We can see that only 1.10\% of the papers consider the EA, RA, SDA, and WHA. These kinds of attacks may cause severe damage to an IoT network.

\subsection{Publisher and Publication Year}\label{sec:publisher}
In this section, we present results regarding the publisher of every paper and the publication year. It would be interesting to see when the research community started to investigate further trust management solutions for the IoT.
\begin{figure}[!t]
\includegraphics[width=\textwidth]{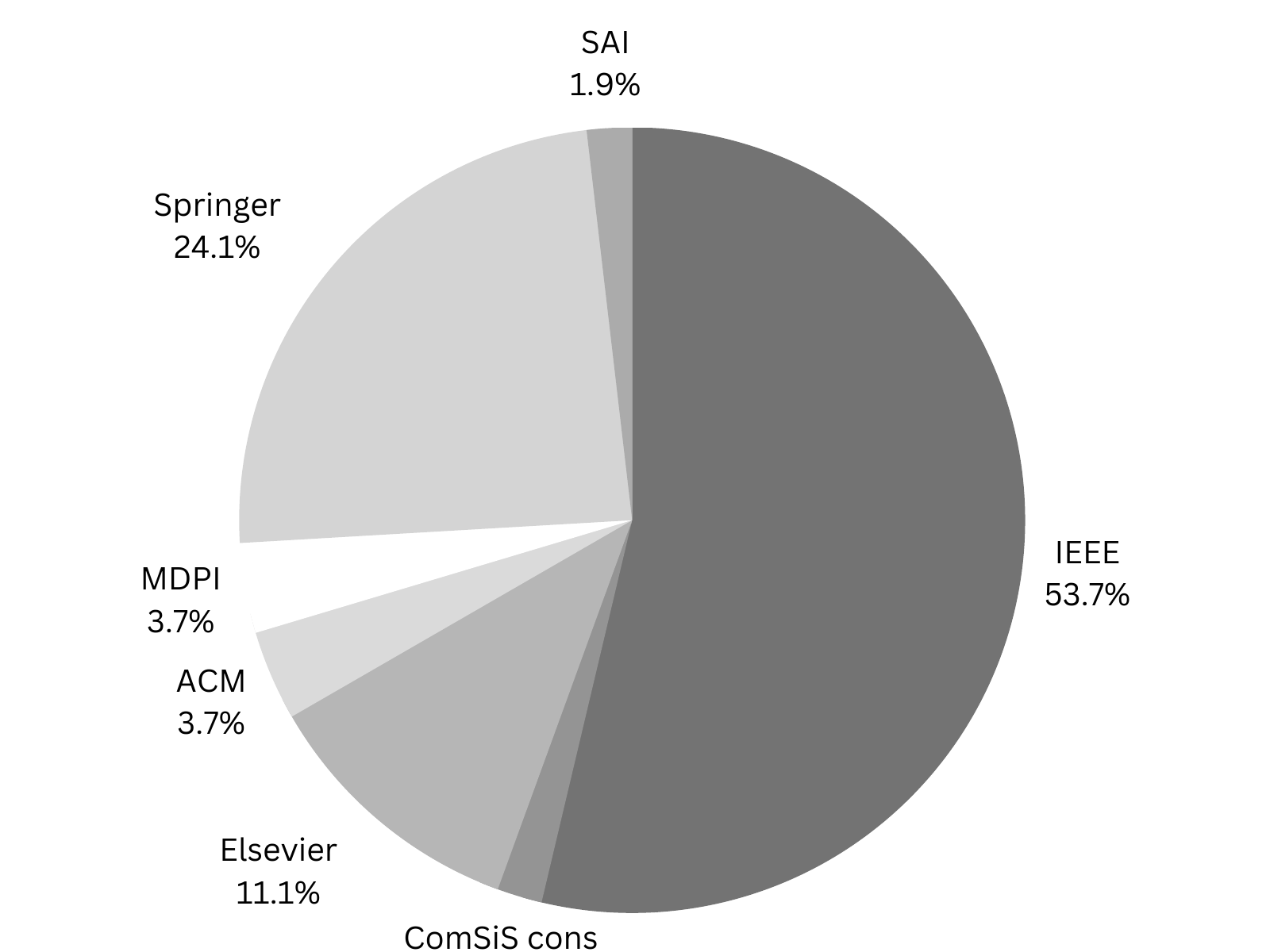}
\caption{Percentage of papers published by each publisher}
\label{fig:pub}
\end{figure}

\begin{figure}[!t]
\includegraphics[width=\textwidth]{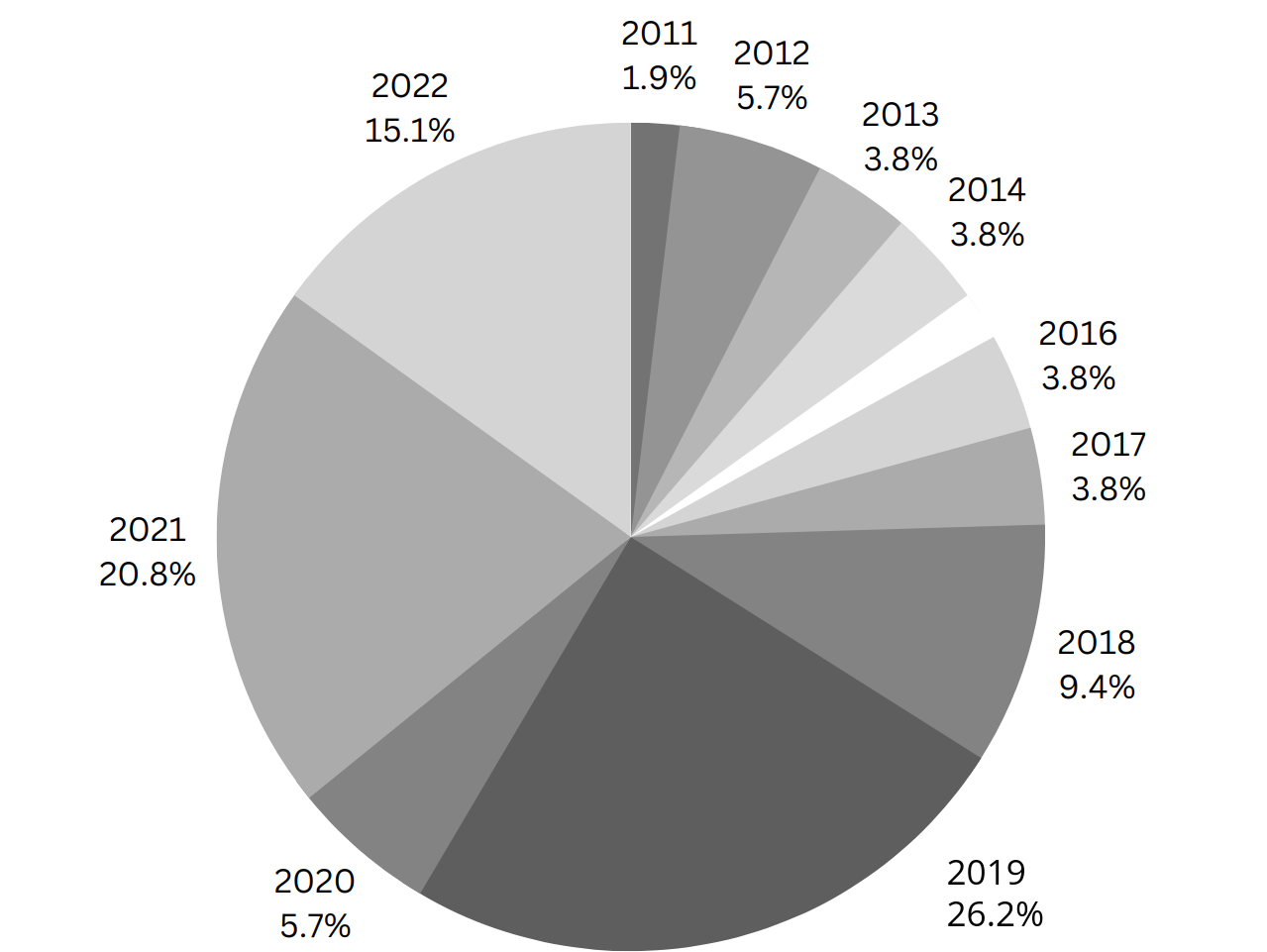}
\caption{Number of papers published in each year}
\label{fig:year}
\end{figure}

We can observe in Figure \ref{fig:pub} that IEEE is the leading publisher, with more than half papers included in this survey. Most of the papers related to trust management for IoT have been published by IEEE.

In Figure \ref{fig:year} we can see that the most popular year for trust management for IoT publications was 2019. The first paper published in Trust Management specifically for IoT was in 2011. It makes sense since IoT is a relatively new technology that has gained a lot of attention in the last decade. The papers involved in this research were published until the end of 2022. 

\subsection{Attacks, Trust Properties, Experiments}\label{sec:attackstrustpro}
In this section, we will give an overview of the categories identified in Table \ref{tab:acategories}. 

\subsubsection{Blockchain}
In the Blockchain category, we included seven papers. Five out of seven papers included a threat model and trust parameters, while all of them included a section with the experiments. 

The attacks that gained the most attention were the trust-related attacks BSA and BMA. Resource-constrained IoT devices are not always capable of participating in a blockchain network as foul nodes, so they are vulnerable to EA. However, EA was studied in one paper. 

During the experiments, only a small percentage of the works took into account scalability, and there was no work that took into account energy consumption. Both of these two components are important, especially for extended IoT networks.

\subsubsection{Context}
In this category, we included six papers. One out of six specified a threat model, five out of six defined the trust parameters, and all of them had a section with the experiments. 

Context is a really important parameter for trust. Some individuals may be trustworthy only under specific circumstances and contexts. However, only one paper in this category specifies the threat model, which includes the SA and BMA. Most of the trust-related attacks are not mentioned. A solid trust management system should be able to confront trust-related attacks.

There are no experiments conducted for scalability and energy consumption. A lot of works are comparing their work with other existing models to justify the efficiency of their solution.

\subsubsection{Social}
In this category, we included 23 papers. Twelve out of 23 specified a threat model, 21 out of 23 defined the trust parameters, and 22 of them had a section with the experiments. 

Trust-related attacks are gaining most of the attention. The experiments are focused on performance and comparison with other models. Energy consumption has taken into account a small percentage of papers, i.e. 5\%.

\subsubsection{Fuzzy}
In this category, we included five papers. One out of five specified a threat model, four out of five defined the trust parameters, and all of them had a section with the experiments. 

Trust-related attacks are only mentioned in the research; other attacks are not mentioned in the threat models presented. Also, most of the works are conducting comparison experiments.

\subsubsection{Game Theory}
In this category, we included four papers. Three out of four specified a threat model, two out of five defined the trust parameters, and all of them had a section with experiments. 

Many game-theoretic approaches aim to reduce the energy consumption in an IoT network. So, the trust parameters and the experiments take into account the energy factor for evaluating trust.

\subsubsection{Probabilistic}
In this category, we included six papers. Three out of six specified a threat model, four out of six defined the trust parameters, and all of them had a section with experiments.

The papers are dealing with trust-related attacks. Two probabilistic approaches take into account Energy as a trust parameter and also in the experiments.

\subsubsection{Prediction}
In this category, we included nine papers. Five out of nine specified a threat model, while all of them defined the trust parameters and had a section with the experiments. 
The experiments do not include scalability, and energy. In this category, some non-trust-related attacks are gaining attention, like SFA, SDA, and DoS.

\section{Classification based on the used Methods}\label{class}

This section provides a classification of all the papers considered in the survey. The classification uses the categories described in Section~\ref{sec:intro-cat}. In each category, we will compare the approaches proposed in the papers using several dimensions. To help the reader navigate each category we provide at the beginning of the corresponding section a table that provides an overview of the considered papers. The tables include one column per dimension:  Information Gathering, Trust Update, Experiments, Centralized, Trust Formation, Threat Model, and Simulator as stated in the Background section. Additionally, the last two columns will state the Limitations and Strengths of each method. More in detail: 

\begin{itemize}
    \item Paper: The paper under examination.
    \item Information Gathering: Filled out with the word Direct or Both, in case the paper is using both Direct and Indirect trust. It will remain empty in case the paper does not state how the information gathering is taking place.
    \item Trust update: Filled out with the word Time-Driven, Even-Driven, or Both, in case the paper uses both approaches. It will remain empty if the paper does not state how the trust update is taking place.
    \item Experiments: If a paper provides experiments supporting the theorems a check mark (\checkmark) will be placed, otherwise it will remain empty.
    \item Centralized: If a central entity is responsible for Trust Management, a check mark (\checkmark) will be placed, otherwise it will remain empty.
    \item Trust Formation: Filled out with the word Single-Trust or Multi-Trust, depending on which approach the paper is using. It will remain empty if the paper does not state how the trust formation is taking place or state the trust parameters.
    \item Threat Model: When a paper explicitly states the threat model, a check mark (\checkmark) will be placed; otherwise, it will remain empty.
    \item Simulator: When a paper states which simulator was used to conduct the experiments, a check mark (\checkmark) will be placed; otherwise, it will remain empty.
    \item Limitations: Summarize the limitations of the paper.
    \item Strengths: Summarize the strengths of the paper.
\end{itemize}

The tables can also be used to navigate through the literature by providing the reader with key characteristics of every work. To further support the reader, Table~\ref{tab:multicolov}, provides the overview of all the papers of the literature and their placement in the categories and dimensions.

\begin{table*}[]\scriptsize
\caption{Overall overview of the categories}
\begin{center}
\resizebox{\textwidth}{!}{\begin{tabular}{|| c | c | c | c | c | c | c | c | c | c | c ||}
 \hline
 Property & Subvalues & \makecell{Direct Trust}   & Recommendations & Fuzzy Logic & Blockchain & Game Theory & Context & Social & Prediction & Probabilistic  \\ [0.5ex] 
 \hline\hline

    \multirow{2}{*}{\makecell{Info \\ Gathering}}& Direct & \makecell{\cite{Alshehri,Dedeoglu,Ma, WangMTES}, \\ \cite{SAIED, Joshi, Subhash, Jayasinghe}} & & \makecell{\cite{Alshehri}} & \cite{Dedeoglu}& & \cite{SAIED}& \cite{Jayasinghe}& \cite{Subhash, Jayasinghe}& \cite{Joshi, WangMTES} \\\cline{2-11}
    & \makecell{Direct \& \\ Indirect}& & \makecell{\cite{Aldawsari2021, QURESHI2020103756, Din, Yu}, \\ \cite{Mendoza, AwanRob, DIN2022108013}}& \makecell{\cite{Guleng, Mahmud, Chen}} & \makecell{\cite{Putra, Kouicem, Amiri-Zarandi, Jeribi}}& \cite{Djedjig, Rani, Duan}& \makecell{\cite{Rafey, ALTAF, Abidi, Adewuyi}, \\ \cite{Magdich1}} &  \makecell{\cite{Rafey, Nitti, Mon, Marche}, \\ \cite{Abidi, NittiSUB, Wang, ChenSOA}, \\ \cite{Bao, Kowshalya, Adewuyi, BaoSCA}, \\ \cite{Das, Awan, Amiri-Zarandi, BaoAPP}, \\ \cite{Chen_trust-basedservice, Wen, Abderrahim, Aalibagi}, \\ \cite{MAGDICH202292, Magdich1}}& \makecell{\cite{Wang, Wen, Alnumay, Abderrahim}, \\ \cite{Aalibagi, Magdich1, MAGDICH202292}}& \cite{Boudagdigue, Wang, Fang, FangFETMS}  \\
    \hline 

    \multirow{3}{*}{\makecell{Trust \\ Update}}& Time-Driven & \makecell{\cite{Joshi}}& \makecell{\cite{Aldawsari2021}}& \makecell{\cite{ESPOSITO}}& \cite{Kouicem, ESPOSITO, Jeribi} & \cite{ESPOSITO, Rani, Duan}& & \cite{Mon, Abderrahim, MAGDICH202292}& \cite{Abderrahim, MAGDICH202292} & \cite{Joshi} \\\cline{2-11}
    & Event-Driven & \makecell{\cite{Dedeoglu, Ma, SAIED}} & \makecell{\cite{QURESHI2020103756, Din, Mendoza, AwanRob}}& \makecell{\cite{Guleng, Chen}} & \makecell{\cite{Dedeoglu, Debe, Putra, Bordel}, \\ \cite{Amiri-Zarandi}}& & \makecell{\cite{Rafey, ALTAF, SAIED, Abidi}, \\ \cite{Adewuyi, Magdich1}}& \makecell{\cite{Rafey, Nitti, Marche, Abidi}, \\ \cite{NittiSUB, ChenSOA, Bao, Kowshalya}, \\ \cite{Adewuyi, BaoSCA, Amiri-Zarandi, BaoAPP}, \\ \cite{Chen_trust-basedservice, Wen, Magdich1}} & \cite{Wen, Alnumay, Magdich1}& \cite{FangFETMS} \\\cline{2-11}
    & Both & & & \makecell{\cite{Mahmud}} & & \cite{Djedjig} & & \cite{Awan}& &  \\
    \hline 

    \multirow{2}{*}{\makecell{Trust \\ Formation}}& Multi-Trust & \makecell{\cite{Alshehri,Dedeoglu, Ma, WangMTES}, \\ \cite{SAIED, Joshi, Subhash, Jayasinghe}} & \makecell{\cite{ QURESHI2020103756, Din, Yu,Mendoza}, \\ \cite{ AwanRob, DIN2022108013}} & \makecell{\cite{Alshehri, Guleng, Mahmud, Chen}}& \makecell{\cite{Dedeoglu, Putra, Kouicem, Bordel}, \\ \cite{Jeribi}} & \cite{Djedjig}& \makecell{\cite{Rafey, ALTAF, SAIED, Abidi}, \\ \cite{Adewuyi, Magdich1}} & \makecell{\cite{Rafey, Nitti, Mon, Marche}, \\ \cite{Abidi, NittiSUB, Wang, Bao}, \\ \cite{Kowshalya, Adewuyi, Das, Awan}, \\ \cite{Jayasinghe, BaoAPP, Chen_trust-basedservice, Wen}, \\ \cite{Abderrahim, Aalibagi, MAGDICH202292, Magdich1}} & \makecell{\cite{Wang, Subhash, Wen, Alnumay}, \\ \cite{Abderrahim, Jayasinghe, Aalibagi, MAGDICH202292}, \cite{Magdich1}}&  \cite{Boudagdigue, Wang, Joshi, WangMTES}\\\cline{2-11}
    & Single-Trust & & \makecell{\cite{Aldawsari2021}} & & \cite{Debe, Amiri-Zarandi}& \cite{Duan}& & \cite{Amiri-Zarandi} & &  \\
    \hline

    Experiments &  - & \makecell{\cite{Alshehri, Dedeoglu, Ma, WangMTES}, \\ \cite{SAIED, Joshi, Subhash, Jayasinghe}} & \makecell{\cite{Aldawsari2021, QURESHI2020103756, Din, Yu}, \\ \cite{Mendoza, AwanRob, DIN2022108013}} & \makecell{\cite{Alshehri, Guleng, Mahmud, Chen} \\ \cite{ESPOSITO}}& \makecell{\cite{Dedeoglu, Debe, Putra, Kouicem} \\ \cite{Bordel, Amiri-Zarandi, ESPOSITO, Jeribi}}& \cite{Djedjig,ESPOSITO, Rani, Duan} & \makecell{\cite{Rafey, ALTAF, SAIED, Abidi}, \\ \cite{Adewuyi, Magdich1}}  &  \makecell{\cite{Rafey, Nitti, Mon, Marche}, \\ \cite{Abidi, NittiSUB, Wang, ChenSOA}, \\ \cite{Bao, Kowshalya, Adewuyi, BaoSCA}, \\ \cite{Das, Jayasinghe, Amiri-Zarandi, BaoAPP}, \\ \cite{Chen_trust-basedservice, Wen, Abderrahim, Aalibagi}, \\ \cite{MAGDICH202292, Magdich1}}& \makecell{\cite{Wang, Subhash, Wen, Alnumay}, \\ \cite{Abderrahim, Jayasinghe, Aalibagi, MAGDICH202292}, \cite{Magdich1}} & \makecell{\cite{Boudagdigue, Wang, Joshi, WangMTES}, \\ \cite{Fang, FangFETMS}}\\

    \hline 

    Centralized &  - & \makecell{\cite{Ma, SAIED}} & \makecell{\cite{Din, DIN2022108013}} & & & & \cite{SAIED} & \cite{Mon, Awan}& & \\

    \hline

    Threat Model &  - & \makecell{\cite{Alshehri,Dedeoglu, Ma, Joshi}} & \makecell{\cite{QURESHI2020103756}}& \makecell{\cite{Alshehri, ESPOSITO}} & \makecell{\cite{Dedeoglu, Debe, Putra, Kouicem} \\ \cite{Bordel, Amiri-Zarandi, ESPOSITO}} & \makecell{\cite{Dedeoglu, Bordel, Jeribi}}& \cite{Rafey}& \makecell{\cite{Rafey, Marche, ChenSOA, Bao}, \\ \cite{BaoSCA, Amiri-Zarandi, BaoAPP, MAGDICH202292}, \\ \cite{Chen_trust-basedservice, Wen, Abderrahim, Aalibagi}} & \makecell{\cite{Wen, Alnumay, Abderrahim, Aalibagi}, \\ \cite{MAGDICH202292}}& \cite{Boudagdigue, Joshi, FangFETMS}\\

    \hline

    Simulator &  - & \makecell{\cite{Alshehri,Dedeoglu, WangMTES, Joshi}, \\ \cite{Subhash}} & \makecell{\cite{Aldawsari2021, QURESHI2020103756, Yu, Mendoza}, \\ \cite{AwanRob}} & \makecell{\cite{Alshehri, Guleng, Mahmud, Chen}} & \makecell{\cite{Dedeoglu, Bordel, Jeribi}} & \cite{Djedjig, Rani}& \cite{Abidi, Magdich1}& \makecell{\cite{Abidi, ChenSOA, Das, Chen_trust-basedservice}, \\ \cite{Abderrahim, Magdich1, MAGDICH202292}}& \makecell{\cite{Subhash, Abderrahim, Aalibagi, MAGDICH202292}, \\ \cite{Magdich1}}& \cite{Joshi, WangMTES, Fang, FangFETMS}\\

    \hline

\end{tabular}}
\end{center}
\label{tab:multicolov}
\end{table*}

\subsection{Direct Trust}
Direct observations refer to the process of gathering information for trust calculation through direct communication between the nodes. The node relies on its own observation when the trust calculation is taking place.

A summary of the works relying only upon direct observations for trust evaluation is presented in Table \ref{tab:direct}. We can observe from the table that all the works provide experiments, prefer the multi-trust approach, and use the direct trust method - as stated by the category. We can see that 62.5\% of the papers mention the simulator used. The Threat model is explicitly defined in 50\% of the papers. We can also observe that 25\% of the papers provide a Centralized approach.  Finally, 37.5\% of the approaches are Event-Driven, while 50\% of the papers do not define the Trust Update procedure. In the rest of the section, we provide a summary of each paper.

Alshehri et al. \cite{Alshehri} proposed a cluster-based architecture, including one super node and many master nodes that are responsible for multiple cluster nodes. Only the cluster nodes are considered malicious. The malicious nodes can perform OOA. Alshehri et al. introduced five algorithms to calculate the trust score of every cluster node. To calculate the trust score, they take into account the quality of service, the history score, and the trust score. They use a fuzzy approach to classify the trust scores into fuzzy sets. The nodes are then classified into 3 categories: trusted, semi-trusted, and non-trusted. Based on the trust score, the nodes can change clusters, and based on the category, they can perform specific acts. They performed experiments in the Cooja simulator regarding scalability, the accuracy of different attacks, and fuzzy and non-fuzzy, approaches and they presented diagrams with the results. The authors do not specify the trusted entities involved in the procedure. Also, they only deal with one trust-related attack. Finally, they proposed a HEXA decimal-based messaging system that can be used to detect tampered messages in transit, and they isolate the untrusted nodes from the network.

Dedeoglu et al. \cite{Dedeoglu} proposed a system containing sensors and gateways. Gateways run the blockchain and are associated with several sensors. A malicious sensor can tamper with the data, and a malicious gateway can generate invalid blocks. A lightweight block generation scheme was proposed where blocks are generated at time intervals. The block validation mechanism adapts the block validation scheme based on the reputation of the node that generated the block and the number of validators. For the consensus mechanism, the following method was introduced: If a validator detects an invalid transaction, it broadcasts INVALID and the nodes have to validate the transaction;  otherwise, the block is appended to the blockchain. The proposed technique evaluates the trustworthiness of sensor observations. The sensor assigns a confident value to the data and sends it to a gateway. The gateway compares the data with the data of the other cluster sensors (assumption: the sensors in the same cluster have correlated data). In the end, based on the result, the reputation of the node is recalculated. The gateways store the information on a blockchain. The reputation of the gateways is calculated based on their actions during the generation and validation of the blocks. The trust parameters taken into account are the confidence of the data source, the reputation of the data source, and evidence from other observations. The experiments took place in the NS-3 simulator, and they are both blockchain-related and trust-related. The final results were presented in diagrams. One limitation of this approach is that the neighboring sensors have to gather the same category of data; otherwise, it cannot be applied. 

Ma et al. \cite{Ma} proposed a multi-mix attack method. The sub-attacks include: tamper, replay, and drop attacks. The nodes update their cognition when a packet is transferred. The base station collects all cognitions from nodes and performs a central trust evaluation. For detection of the malicious nodes, the node’s trust should be forwarded to the k-means clustering module. The trust properties used for trust evaluation are honesty, straightness, and volume. During the experiments, the accuracy of the proposed method was tested. This paper does not deal with trust-related attacks, but it considers a type of attacker that can perform mixed attacks.

Wang et al. \cite{WangMTES} proposed a system consisting of Mobile edge nodes (MEN) and common sensors. The MEN are connected to a small number of sensors. In this paper, a mobile edge trust evaluation scheme is proposed. The evaluation of the trustworthiness of sensor nodes is achieved using a probabilistic graph model. The probabilistic graph model is used to represent the relationship between nodes. The interaction of node $i$ with node $j$ can be described as $P$ and $Q$. $P$ is a positive influence of node $i$ on node $j$ and $Q$ is a negative one. The information gathered for the formation of trust is the result of data collection and communication behavior. Also, a moving strategy method is proposed to decrease the travel distance MEN has to cover to evaluate every sensor. The experiments were conducted using MATLAB and NS-3 and were focused on the performance of the mechanism, the analysis of energy consumption, and the testing of the proposed moving algorithm. The results were presented in diagrams. The paper does not specify the threat model, which is a drawback. A strength, on the other hand, is that the authors propose a moving strategy for energy savings in a high-mobility environment.

Saied et al. \cite{SAIED} proposed a context-aware and multi-service trust management system. Upon a request from a node asking for assistance, the trust manager starts the entity selection process to return a set of trustworthy assisting nodes to the requester. A set of recommenders sends reports; the most important are those that lie to the same or more similar services and recent ones. A quality of recommendation score is assigned to each node reflecting its trustworthiness when rating other nodes. The context was used to filter out recommendations and select the most relevant ones. The trust is calculated based on the following parameters:
The score is given by the requester node to the service provider evaluating the offered service, a weight that depends on time similarity, and quality of recommendations. Experiments were performed focused on the comparison of reactions against different kinds of attacks like on-off, bad-mouthing, and selective behavior attacks. The authors do not specify the threat model, which is a drawback. On the other hand, they involve the context in the trust-related procedure. A node can act differently in different contexts.

Joshi et al. \cite{Joshi} proposed a system that consists of several resource-constrained IoT nodes with a short radio range and a base station with a limitless source of energy as a central authority. This research work has presented a 2-state HMM with a Trusted state and a compromised state, together with essential and unessential output as observation states. The trustworthiness of the node is modeled by the 2-state HMM to predict the likelihood of the node's next state. The state transition probability matrix is defined by the energy consumed, the number of modified packets, and the number of forwarding packets. The malicious nodes can drop the packets or tamper with the data. Experiments were conducted in MATLAB to evaluate the network's trustworthiness with various percentages of compromised nodes and compare it with other methods. The results were presented in diagrams. The authors are taking into account only two kinds of attacks. The authors are using energy consumption as a key characteristic for calculating trust. This is interesting since increased activity might be malicious, but also energy of the nodes is also taken into account in a resource-constrained environment.

Subhash et al. \cite{Subhash} proposed the Power Trust. Power Trust assigns trust values to the nodes of the network based on energy auditing. Using the energy auditing model, they calculate the trust values of every node present in the network dynamically and predict physical and cyberattacks. To detect the attacks, a deep learning model was trained with past data that contains normal and excessive energy consumption due to an attack. The model can predict both physical and cyberattacks. The experiments were performed in the Cooja simulator, and they were focused on the performance and the accuracy of the method. The results were presented in diagrams. The authors do not state the threat model, and they do not give details about the deep learning model used. The method takes into account energy consumption, which is important in a resource-constrained environment. Finally, they also predict both physical and cyberattacks.

Jayasinghe et al. \cite{Jayasinghe} proposed a system where the nodes form communities of interest. In this model, the transactions are under evaluation, and should be determined if a transaction is trustworthy. The trust parameters used are co-location relationships, co-work relationships, mutuality and centrality, and cooperativeness. An unsupervised learning technique was employed to label the data's trustworthiness. After the labeling, an SVM model predicts the trust level. Experiments were performed to observe the performance of the proposed solution. The authors do not specify the threat model. They offer a metric that provides a perception of a node before interaction. This can be used when there have been no previous interactions or a new node has just entered the network.

\begin{table*}[!t]\scriptsize
  \caption{Overview of approaches proposing direct trust Trust Management methods}
  \label{tab:direct}
  \centering
\resizebox{\textwidth}{!}{\begin{tabular}{||c | c | c | c | c | c | c | c | c | c||} 
 \hline
 Paper & \makecell{Info \\ Gathering}   & Trust Update & Experiments & Centralized & \makecell{Trust \\ Formation} & \makecell{Threat \\ Model} & Simulator & Limitations & Strengths  \\ [0.5ex] 
 \hline\hline
 \cite{Alshehri} & Direct  &  & \checkmark &  &Multi-Trust & \checkmark& \checkmark& \makecell[l]{$\bullet$ Does not specify the trusted entities \\ $\bullet$ Deals only with one trust-related attack.} &  \makecell[l]{$\bullet$ Messaging system for identifying tampered messages \\ $\bullet$ Isolated untrusted nodes}\\ 
 \hline
 \cite{Dedeoglu} & Direct  & Event-Driven & \checkmark &  & Multi-Trust & \checkmark & \checkmark& \makecell[l]{$\bullet$ Neighbor sensors have to gather the same category of data.} & \makecell[l]{$\bullet$ Customized blockchain for IoT} \\
 \hline
 \cite{Ma} & Direct  & Event-Driven  & \checkmark& \checkmark& Multi-Trust & \checkmark& & \makecell[l]{$\bullet$ Do not deal with trust-related attacks} & \makecell[l]{$\bullet$ Considering mixed-attacks} \\  
 \hline
  \cite{WangMTES} & Direct    &   & \checkmark & & Multi-Trust &  & \checkmark & \makecell[l]{$\bullet$ Do not specify the threat model} &\makecell[l]{$\bullet$ Travel strategy for energy saving \\ $\bullet$ Mobility} \\
 \hline
  \cite{SAIED} & Direct   &  Event-Driven & \checkmark & \checkmark& Multi-Trust &  &  & \makecell[l]{$\bullet$ Do not specify the threat model} & \makecell[l]{$\bullet$ Takes into account the context} \\
  \hline
   \cite{Joshi} & Direct   & Time-Driven & \checkmark & & Multi-Trust & \checkmark & \checkmark &  \makecell[l]{$\bullet$ Deals with only 2 kind of attacks} &  \makecell[l]{$\bullet$ Takes into account the energy consumption}\\
 \hline
 \cite{Subhash} & Direct  &  & \checkmark& & Multi-Trust &  & \checkmark & \makecell[l]{$\bullet$ Do not specify the threat model \\ $\bullet$ Do not give details about the deep learning}& \makecell[l]{$\bullet$ Predict the physical attacks and cyber-attacks \\ $\bullet$ Focused on energy consumption}\\
 \hline
  \cite{Jayasinghe} & Direct &  &\checkmark & & Multi-Trust &  & & \makecell[l]{$\bullet$ Do not specify the threat model} & \makecell[l]{$\bullet$ Metric that provides a perception \\ of a node before interaction}\\[1ex] 
 \hline
 \end{tabular}}

\end{table*}

\subsection{Recommendations}
To gather the information needed for the trust calculations, the nodes can ask for recommendations concerning the node under evaluation from other nodes. This procedure is also called indirect trust. There are many reasons why recommendations are valuable for trust evaluation. Some works use recommendations as a supplement to direct observations; the summary of these works is presented in Table~\ref{tab:indirect}. In this category, we include works that only use Recommendations as a method. Works in this category do not fall into other categories. 

From the table, we can observe that all of the works use both direct and indirect trust - as stated by the category. Regarding the Trust Update procedure, most of the papers are using the Event-Driven. More specifically, 57.14\% are Event-Driven, 14.3\% are using the Time-Driven approach and the rest do not refer to the Trust Update Procedure. All papers present experiments, and 71.4\% present the simulator used for the experiments. Moreover, 28.6\% preferred a Centralized approach. Regarding Trust Formation, most of the papers present a Multi Trust approach (85.7\%) and the rest of the papers provide a Single Trust approach. Finally, 14.3\% of the papers define the Threat Model. In the rest of this part, we are going to present a summary of each work.

Aldawsari et al. \cite{Aldawsari2021} proposed a cluster-based system.  They also incorporated a base station (BS) with unlimited energy into the network. For evaluating trust at the cluster level the direct trust of the Cluster-Head and the recommendations of its neighbors are encountered. For trust between
two different clusters the cluster heads and the
BS is participating in the procedure. 
The energy consumption is taken into account to calculate the trust. Experiments were conducted on the NS-3 simulator to test the detection rate, energy consumption, and trust evaluation time and to compare the proposed scheme with other methods. The results were presented in diagrams. The authors do not specify the threat model, which is a limitation. It is important that energy consumption be taken into account in a resource-constrained environment.

Qureshi et al. \cite{QURESHI2020103756} proposed a trust management system for edge-based IoT networks. The proposed model combines direct and indirect trust to derive the trust level. The system's threat model includes BMA, DoS, and OOA. The trust calculation procedure takes into account the packet drop rate and the packet data rate. OMNET++ used for the following experiments: level of trustworthiness, detection rate, detection accuracy, detection of false positive rate, impact of a network lifetime, impact of average packet delay, the impact of average throughput, and end-to-end delay analysis. As a limitation, we state that the threat model can be expanded, so the method will be able to identify more attacks. The authors are considering an edge-based IoT architecture, which is important since IoT and edge devices are collaborating on multiple concepts.

Din et al. \cite{Din} proposed a mechanism consisting of IoT-edge nodes, an application programming interface, and a centralized trust agent. The trust agent evaluates the trust level. Based on their trust level, the nodes are allowed to communicate with other nodes. The trust properties used are the following:  compatibility, cooperativeness, delivery ratio, and recommendations. The Contiki Cooja simulator was used to acquire the results, the Java language was used for the interaction, and a virtual machine was a platform for simulations. The experiments focused on testing the quality of service (QoS) and resilience
against the BMA, BSA, WHA, and SPA. Even though the authors are presenting experiments for specific attacks, in the main part of the paper they do not specify the threat model.

Yu et al. \cite{Yu} proposed a system consisting of nodes and a base station (BS). The trust parameters for direct trust are the packet forwarding capacity, the repetition rate, the consistency of the packet content, the delay, the integrity, etc. For the calculation of indirect trust, the D-S theory was used. Experiments were conducted on MATAB to test the performance of the solution and its behavior under different attacks and compare the method to other schemes. Also, energy consumption was tested, a crucial parameter in a resource-constrained environment. The authors do not specify the threat model.

Mendoza et al. \cite{Mendoza} proposed a distributed trust management model for multi-service IoT using
direct and indirect observations. The trust management scheme assigns positive scores for honest nodes and
negative scores for malicious nodes, using direct interactions between nodes (service requests)
and recommendations from neighbors (by exchanging trust tables). The authors implemented malicious nodes performing the BMA to analyze the effectiveness of the model.
The obtained results from the experiments conducted on the Cooja simulator show the proposed trust management model detects malicious behavior in the network
considering topologies with 10\% to 30\% of malicious nodes. This model may be used to
detect other common attacks in the IoT. The authors do not specify the threat model.

Awan et al. \cite{AwanRob} proposed a mechanism that mainly works with direct observations and asks for recommendations if no interactions were recorded in the past. The authors also took into account scalability, since the nodes only store the results of the experience component. Other trust parameters are reputation and knowledge. The following experiments took place on the NS-3 simulator: the behavior of the solution was tested on BSA, BMA, and OOA; comparison with other models; and energy consumption. Despite the fact that the authors are conducting experiments under different attacks, they do not specify the threat model in the main part of the paper or the architecture of the network. The nodes are storing one component of trust for scalability and storage reasons and the authors are taking into account the energy consumption.

Din et al. \cite{DIN2022108013} are considering a resource-sharing environment consisting of resource providers and resource seekers. The network also has a central authority that coordinates the procedures. The resource providers are nodes that want to share their resources, and the resource seekers are in need of more resources. A resource is allocated to a specific node for a specific amount of time. The resource providers have to make an offer to the interface, and the resource seekers check the availability and choose the offer that best fits their needs. When a resource seeker wants to use some resources from the resource provider, the trust evaluation procedure is triggered using previous observations. The trust evaluation consists of two components: competence, which involves stability and cooperativeness, and trustworthiness, which includes persistence and reputation. During the first interaction, the procedure takes into account only recommendations. The final trust values are compared to a threshold. The authors do not specify the threat model, but they present a framework where resource-constrained nodes can share and exploit free resources from other nodes in the network.

\begin{table*}[!t] \scriptsize
  \caption{Overview of approaches proposing Trust Management methods using Recommendations}
  \label{tab:indirect}
  \centering
\resizebox{\textwidth}{!}{\begin{tabular}{||c | c | c | c | c | c | c | c | c | c||} 
 \hline
 Paper & \makecell{Info \\ Gathering}   & Trust Update & Experiments & Centralized & \makecell{Trust \\ Formation} & \makecell{Threat \\ Model} & Simulator & Limitations & Strengths  \\ [0.5ex] 
 \hline\hline
   \cite{Aldawsari2021} & Both &  Time-Driven & \checkmark & & Single-Trust &  & \checkmark & \makecell[l]{$\bullet$ Do not specify the threat model}& \makecell[l]{$\bullet$ Takes into account energy consumption}\\
 \hline
 \cite{QURESHI2020103756} & Both &  Event-Driven & \checkmark & & Multi-Trust & \checkmark  & \checkmark & \makecell[l]{$\bullet$ Limited attacks in threat model}& \makecell[l]{$\bullet$ Edge-based IoT} \\ 
 \hline
  \cite{Din} & Both &  Event-Driven & \checkmark & \checkmark & Multi-Trust &  & & \makecell[l]{$\bullet$ Do not specify the threat model}& \makecell[l]{$\bullet$}\\ 
 \hline
   \cite{Yu} & Both &   & \checkmark &  & Multi-Trust &  & \checkmark & \makecell[l]{$\bullet$ Do not specify the threat model}& \makecell[l]{$\bullet$ Energy consumption was tested}\\ 
 \hline
   \cite{Mendoza} & Both &  Event-Driven & \checkmark & & Multi-Trust &   & \checkmark & \makecell[l]{$\bullet$ Do not specify the threat model}& \makecell[l]{$\bullet$}\\ 
 \hline
  \cite{AwanRob} & Both &  Event-Driven & \checkmark & & Multi-Trust &   & \checkmark & \makecell[l]{$\bullet$ Do not specify the threat model \\ $\bullet$ Do not specify the architecture of the network}& \makecell[l]{$\bullet$ The nodes only store one component of trust for scalability \\ $\bullet$ Takes into account energy consumption}\\
  \hline
  \cite{DIN2022108013} & Both &   & \checkmark & \checkmark & Multi-Trust &   &  & \makecell[l]{$\bullet$ Do not specify the threat model}& \makecell[l]{$\bullet$ Focused on resource sharing}\\ [1ex]
 \hline
 \end{tabular}}

\end{table*}

\subsection{Fuzzy Logic} Boolean logic permits expressions that are either true or false. However, in real life sometimes truth is a spectrum. Fuzzy logic is a multi-value logic that permits intermediate values between true and false \cite{Castillo2001}\cite{Hurlimann2009}. Some works on trust management exploit the nature of fuzzy logic to represent trust as a spectrum between trusted and not trusted. A summary of the findings is presented in Table \ref{tab:fuzzy}.

We can observe from the table that all of the papers provide experiments. We can see that 80\% of the papers mention the simulator used to conduct the experiments. Regarding the Trust Update, 60\% of the paper preferred an Event-Driven approach, while the rest (40\%) the Time-Driven method. It is interesting to notice that none of the papers use a Centralized authority for trust evaluation, and none of the papers prefer a Single Trust approach. It is also worth pointing out that 20\% of the papers in this category do not use direct or indirect trust. Finally, 40\% of the papers define the Threat Model.

Alshehri et al. \cite{Alshehri} proposed a cluster-based architecture, including one super node and many master nodes that are responsible for multiple cluster nodes. Only the cluster nodes are considered malicious. The malicious nodes can perform OOA. Alshehri et al. introduced five algorithms to calculate the trust score of every cluster node. To calculate the trust score, they take into account the quality of service, the history score, and the trust score. They use a fuzzy approach to classify the trust scores into fuzzy sets. The nodes are then classified into three categories: trusted, semi-trusted, and non-trusted. Based on the trust score, the nodes can change clusters, and based on the category, they can perform specific acts. They performed experiments in the Cooja simulator regarding scalability, the accuracy of different attacks, and fuzzy and non-fuzzy approaches, and they presented diagrams with the results.  The authors do not specify the trusted entities involved in the procedure. Also, they only deal with one trust-related attack. Finally, they proposed a HEXA decimal-based messaging system that can be used to detect tampered messages in transit, and they isolate the untrusted nodes from the network.

Guleng et al. \cite{Guleng} proposed a trust management architecture for vehicular ad hoc networks (VANET). Consequently, we can conclude that they studied dynamic topology. The proposed scheme is distributed and uses fuzzy logic to evaluate direct trust. For indirect trust, the authors proposed a reinforcement learning approach. To calculate the trust scores, they take into account the cooperativeness, honesty, responsibility factor, and previous values. Also, the proposed trust management takes into account the trust in a message. They performed experiments using the NS-2.34 simulator to compare their method with "w/o Trust" and "Deterministic Trust". Also, a simulation of BMA was performed. The final results of the experiments were presented in diagrams. The proposed solution does not specify the threat model. It is a nice approach that the authors are taking into account both the node's trust and the message's trust. The solution can also be applied to a high-mobility network.

Mahmud et al. \cite{Mahmud} proposed a trust management method for cloud-based architecture for neuroscience applications. The proposed method estimates the trust level using an adaptive neuro-fuzzy inference system (ANFIS) and weighted additive methods. Furthermore, it is worth pointing out that this technique takes into account the behavior of the node and the trustworthiness of the generated data. Behavioral trust takes into account: the Relative Frequency of Interaction, intimacy, honesty, previous interactions, and indirect trust. Data trust depends on the deviation of a node's instantaneous data from its historical data and indirect recommendations. NS-2 simulator was used to perform the following experiments: Packet Forwarding Ratio, Network Throughput, Average Energy Consumption Ratio, Accuracy, F-measure, Comparison with other models, and different linguistic terms (5 and 3). The final results of the experiments were presented in diagrams. The authors do not specify the threat model. In our opinion, it is a pro that they take into account data's trust and energy consumption.

Chen et al. \cite{Chen} focuses on a dynamic architecture, which means that some nodes may leave or enter the network. The nodes are divided into Service Providers and Service Requestors. This work proposes a distributed fuzzy logic trust management scheme. The model consists of both direct trust (monitoring the neighbors) and indirect trust (recommendations). To calculate the trust, the following values are considered: the end-to-end forwarding ratio (EPFR), the average energy consumption (AEC), and the packet delivery ratio(PDR). Also, a global trust can be issued to obtain a more accurate value of trust. For the experiments, the NS-3 simulator was used, and the following values were taken into account: EPFR, AEC, PDR, convergence speed, detection probability, and comparison with other models. The final results of the experiments were presented in diagrams. There is a need for more efficient global trust computations. 

Esposito et al. \cite{ESPOSITO} proposed a system consisting of IoT nodes that communicate with edge nodes that participate in a blockchain. The blockchain stores a smart contract that periodically stores the trust scores. The smart contract receives the real numbers extracted from the nodes and uses fuzzy logic to translate the real numbers into linguistic terms.
The main threat to the system is to store on the blockchain a false value for the computed trust score. So, the solution is to focus on finding a way to reject these kinds of false messages. A game-theoretic approach was employed, forming a game between the edge node and a common node. A node has two available actions: to send a message or to avoid sending a message. This message might contain malicious data. At the reception of the message, the edge node can do only two
possible actions: Y indicating that it accepts the message and passes it to the blockchain participants to update its state, or N indicating that it rejects the message and does not pass it to the blockchain. Real sensors were used for performing the following experiments: comparison with other models, belief evolution, and attack success probability with and without the proposed defense. The final results of the experiments were presented in diagrams. The authors do not deal with blockchain-related attacks. Messages coming from untrusted sources are blocked.

\begin{table*}[!t]\scriptsize
  \caption{Overview of approaches proposing Fuzzy Logic Trust Management methods}
  \label{tab:fuzzy}
  \centering
\resizebox{\textwidth}{!}{\begin{tabular}{||c | c | c | c | c | c | c | c | c | c||} 
 \hline
 Paper & \makecell{Info \\ Gathering}   & Trust Update & Experiments & Centralized & \makecell{Trust \\ Formation} & \makecell{Threat \\ Model} & Simulator & Limitations & Strengths  \\ [0.5ex] 
 \hline\hline
 \cite{Alshehri} & Direct  &  & \checkmark &  &Multi-Trust & \checkmark& \checkmark& \makecell[l]{$\bullet$ Does not specify the trusted entities \\ $\bullet$ Only one trust-related attack} &  \makecell[l]{$\bullet$ Messaging system for identifying tampered messages \\ $\bullet$ Isolated untrusted nodes}\\ 
 \hline
\cite{Guleng} & Both & Event-Driven & \checkmark & &Multi-Trust & &\checkmark & \makecell[l]{$\bullet$ Do not specify the threat model} & \makecell[l]{$\bullet$ Takes into account both the trust of the node and the message \\ $\bullet$ All nodes can be malicious \\ $\bullet$ Takes into account the mobility}\\
 \hline
 \cite{Mahmud} & Both & Both &\checkmark  & & Multi-Trust& & \checkmark& \makecell[l]{$\bullet$ Do not clearly specify the trust model} & \makecell[l]{$\bullet$ Takes into account the data trust \\ $\bullet$ Takes into account Energy consumption} \\  
 \hline
 \cite{Chen} &  Both & Event-Driven & \checkmark & & Multi-Trust&  & \checkmark& \makecell[l]{$\bullet$ Need for more efficient global trust computation \\ $\bullet$ Do not specify the threat model}& \makecell[l]{$\bullet$ Takes into account Energy consumption} \\ 
 \hline
 \cite{ESPOSITO} &  &  Time-Driven & \checkmark & & & \checkmark&  & \makecell[l]{$\bullet$ Do not study blockchain-related attacks}& \makecell[l]{$\bullet$ Blocks messages from untrusted nodes.} \\[1ex] 
 \hline
 \end{tabular}}

\end{table*}

\subsection{Blockchain Based} 
The blockchain is a distributed database or ledger first introduced in \cite{Nakamoto} as an underlying technology for Bitcoin. Blockchain provides immutability since all the participants are managing the chain through a consensus mechanism. This property made the blockchain popular in different applications. In the scope of trust management for IoT, some works exploit the properties of the blockchain to calculate or store trust data. A summary of the results can be seen in Table \ref{tab:blockchain}.

We can observe from the table that all of the papers provide experiments, but only 37.5\% of the papers provide the simulator used. None of the papers proposes a Centralized approach, which makes sense since we are examining blockchain-based methods. Regarding information gathering: 12.5\% are using only direct trust, 50\% are using recommendations, and 37.5\% do not use an information gathering technique. Most of the approaches are Event-Driven (62.5\%), while the rest (37.5\%) are Time-Driven. In total, 87.5\% are referring to the trust formation, with 71.4\% using the Multi-Trust scheme. Finally, 75\% defines the Threat Model.

Dedeoglu et al. \cite{Dedeoglu} proposed a system containing sensors and gateways. Gateways run the blockchain and are associated with several sensors. A malicious sensor can tamper with the data, and a malicious gateway can generate invalid blocks. A lightweight block generation scheme was proposed where blocks are generated at time intervals. The block validation mechanism adapts the block validation scheme based on the reputation of the node that generated the block and the number of validators. For the consensus mechanism, the following method was introduced: If a validator detects an invalid transaction, it broadcasts INVALID and the nodes have to validate the transaction;  otherwise, the block is appended to the blockchain. The proposed technique evaluates the trustworthiness of sensor observations. The sensor assigns a confident value to the data and sends it to a gateway. The gateway compares the data with the data of the other cluster sensors (assumption: the sensors in the same cluster have correlated data). In the end, based on the result, the reputation of the node is recalculated. The gateways store the information on a blockchain. The reputation of the gateways is calculated based on their actions during the generation and validation of the blocks. The trust parameters taken into account are the confidence of the data source, the reputation of the data source, and evidence from other observations. The experiments took place in the NS-3 simulator, and they are both blockchain-related and trust-related. The final results were presented in diagrams. One limitation of this approach is that the neighboring sensors have to gather the same category of data; otherwise, it cannot be applied. 

Debe et al. \cite{Debe} proposed a scheme based on the Ethereum blockchain where gateways are presented as full nodes of the blockchain and sensors as lightweight nodes. They propose a data attestation solution. The lightweight nodes get some responses from the full nodes. These responses are validated by other full nodes. The attack this work tackles is EA. The trust is calculated on a smart contract, and the parameter used to assess the trustworthiness of a gateway is the client's feedback. The code is publicly available, and the experiments focused on smart contract code vulnerability analysis with tools. This approach stores one trust score per node. If a node is behaving maliciously only toward a small set of nodes, its trust score will still be high. The authors performed cost analysis in terms of gas.

Putra et al. \cite{Putra} proposed a system where sensors are divided into Service Providers and Service Consumers, which consist of the lightweight clients in the main blockchain. Also, a set of permissioned blockchains are implemented to maintain the sensitive data of the sensors. These chains are maintained by a consortium of independent and partially trusted entities. The following attacks can be performed by a malicious node: BMA, RA, PA, BSA, WA, and SA. The trust is calculated on a smart contract on the main blockchain. The experiments took place in a real environment. For the blockchains, the Rinkeby Ethereum test network was used as the main blockchain and private chains were used for storing sensitive data. The following experiments took place: different kinds of weights for calculating the trust scores, comparison with other schemes trust and reputation convergence, latency, and required gas. The final results were presented in diagrams and tables. This approach requires a cluster of partially trusted computers to run the private blockchain. It might be optimistic for some cases to hold such a cluster. On the other hand, this approach must separate sensitive from publicly available data, preserving privacy.

Kouicem et al. \cite{Kouicem} proposed a system based on a fog architecture that consists of the following components: IoT service requesters, Service providers, and fog nodes that are responsible for trust management. The Service providers and the nodes participate in the blockchain. The paper proposes a new consensus mechanism. Each IoT object can assess the trustworthiness of a service provider and share it. Exploiting the blockchain architecture, this protocol provides a global image of trust values. The malicious service providers can perform BMA, SPA, BSA, OOA, and OSA. The malicious fog nodes can drop, delay, modify, and redirect the received messages. The following parameters are taken into account for the trust evaluation procedure: A set of criteria reported on the blockchain for direct trust, previous interactions, and recommendations. For the experiments, a private blockchain was used, and the consensus mechanism was a combination of PBFT and PoS. The following experiments were performed: different kinds of weights, blockchain scalability evaluation, and comparison with other schemes. The results were presented in diagrams. The authors are dealing with only trust-related attacks and make assumptions about the security of the blockchain. However, they deal with a high-mobility environment and propose a new consensus mechanism.

Bordel et al. \cite{Bordel} proposed a method where a trusted third party acts as the TTP. The IoT messages are controlled by a third party. This party is acting between the IoT nodes and the blockchain. The architecture is based on Blockchain technology and the computation of different conceptual models (cognitive, computational, neurological, and game theoretical) using stochastic functions. Smart contracts are employed to calculate global trust. Matlab 2020 was used to perform the following experiments: convergence time and success rate. The results were presented in diagrams. The authors do not specify the threat model or give details about the blockchain technology used. However, they propose multiple concepts of trust.

Amiri-Zarandi et al. \cite{Amiri-Zarandi} proposed a fog architecture scheme. The interactions between the nodes and the blockchain can occur directly or via edge nodes. The nodes are divided into clusters, and they communicate with a fog device. The social connection between the devices is used for trust evaluation. The scheme also works with recommendations that are filtered by a lightweight algorithm. The blockchain is used for storing trust-related data. The malicious nodes can perform BMA, BSA, SFA, and DoS. Honesty was used as the trust parameter. Also, the following experiments took place using the Ethereum blockchain: performance evaluation and experimental comparison with other models. The results were presented in diagrams and tables. The authors do not study any blockchain-related attacks. This approach dynamically selects counselors, as one node might be a good fit at first but behave maliciously afterward. Also, they perform cost analysis in terms of gas.

Esposito et al. \cite{ESPOSITO} proposed a system consisting of IoT nodes that communicate with edge nodes that participate in a blockchain. The blockchain stores a smart contract that periodically stores the trust scores. The smart contract receives the real numbers extracted from the nodes and uses fuzzy logic to translate the real numbers into linguistic terms.
The main threat to the system is to store on the blockchain a false value for the computed trust score. So, the solution is to focus on finding a way to reject these kinds of false messages. A game-theoretic approach was employed, forming a game between the edge node and a common node. A node has two available actions: to send a message or to avoid sending a message. This message might contain malicious data. At the reception of the message, the edge node can do only two
possible actions: Y indicating that it accepts the message and passes it to the blockchain participants to update its state, or N indicating that it rejects the message and does not pass it to the blockchain. Real sensors were used for performing the following experiments: comparison with other models, belief evolution, and attack success probability with and without the proposed defense. The final results of the experiments were presented in diagrams. The authors do not deal with blockchain-related attacks. Messages coming from untrusted sources are blocked.

Jeribi et al. \cite{Jeribi} proposed a solution that includes a network of smart buildings where a lot of IoT devices are installed. There is also a verification manager making access decisions. Each IoT device is connected to a trust management system which is responsible for evaluating the trust of other nodes and producing a complete trust level. The technique takes into account both direct and indirect trust. For direct trust computations, cooperativeness, knowledge, and a group of interest is taken into account. After trust computation, a machine learning algorithm is deployed to classify the trust and determine the most trustworthy device. For this purpose, the ID3 algorithm was used. This algorithm is a supervised learning technique that chooses the best feature that produces the lowest amount of entropy. As an input, it takes an array of trust values and the output is a decision tree of nodes based on the trust values. The trust value of the root node serves as the threshold. Afterward, the values are sent to the blockchain, where they are stored. A permission-based private blockchain was employed. When a new node enters the network, the trust manager calculates the trust value. If it is above the threshold value, it is passed to the blockchain, where it validates that the trust value matches the threshold and is stored in the trustworthy devices. The procedure is repeated at fixed intervals. The method was tested in BSA, BMA, and OOA. This approach requires a network of trusted nodes to run the blockchain, which is not always applicable in a real-life setting. Also, they do not specify the threat model, despite the fact that they perform experiments on specific attacks. This approach deals with the cold-start problem, which is a crucial issue in dynamic environments.

\begin{table*}[!t] \scriptsize
  \caption{Overview of approaches proposing Blockchain-based Trust Management methods}
  \label{tab:blockchain}
  \centering
\resizebox{\textwidth}{!}{\begin{tabular}{||c | c | c | c | c | c | c | c | c | c||} 
 \hline
 Paper & \makecell{Info \\ Gathering}   & Trust Update & Experiments & Centralized & \makecell{Trust \\ Formation} & \makecell{Threat \\ Model} & Simulator & Limitations & Strengths  \\ [0.5ex] 
 \hline\hline
 \cite{Dedeoglu} & Direct  & Event-Driven & \checkmark &  & Multi-Trust & \checkmark & \checkmark& \makecell[l]{$\bullet$ Neighbor sensors have to gather the same category of data.} & \makecell[l]{$\bullet$ Customized blockchain for IoT} \\
 \hline
 \cite{Debe} &  & Event-Driven & \checkmark& & Single-Trust & \checkmark & & \makecell[l]{$\bullet$ One score per node, might be malice to specific nodes \\ $\bullet$ Deals only with one attack} & \makecell[l]{$\bullet$ Available Code \\ $\bullet$ Cost analysis in terms of gas} \\
 \hline
 \cite{Putra} & Both & Event-Driven  & \checkmark& & Multi-Trust & \checkmark &  & \makecell[l]{$\bullet$ Many partially trusted computers to run the private chains \\ $\bullet$ Issue with a new node with 0 score \\ $\bullet$ Only trust-related attacks} & \makecell[l]{$\bullet$ Separation between sensitive and publicly available data.} \\
 \hline
  \cite{Kouicem} & Both & Time-Driven  &\checkmark & & Multi-Trust &  \checkmark &  & \makecell[l]{$\bullet$ Make assumptions about blockchain-related attacks \\ $\bullet$ Only trust-related attacks}& \makecell[l]{$\bullet$ Mobility \\ $\bullet$ New consensus method}\\
 \hline
   \cite{Bordel} &  &  Event-Driven &\checkmark  & & Multi-Trust &  & \checkmark & \makecell[l]{$\bullet$ Do not specify the threat model \\ $\bullet$ Do not give details about the blockchain technology used} &\makecell[l]{$\bullet$ Multiple trust concepts} \\
 \hline
  \cite{Amiri-Zarandi} & Both & Event-Driven & \checkmark& & Single-Trust & \checkmark & & \makecell[l]{$\bullet$ Do not study blockchain-related attacks} & \makecell[l]{$\bullet$ Dynamically selected counselors \\ $\bullet$ Cost analysis in terms of gas} \\ 
  \hline
 \cite{ESPOSITO} &  &  Time-Driven & \checkmark & & & \checkmark&  & \makecell[l]{$\bullet$ Do not study blockchain-related attacks}& \makecell[l]{$\bullet$ Blocks messages from untrusted nodes.} \\
 \hline
 \cite{Jeribi} & Both &  Time-Driven & \checkmark & & Multi-Trust & & \checkmark & \makecell[l]{$\bullet$ A trusted network of nodes is required to run the blockchain \\ $\bullet$ Focused on smart buildings \\ Do not specify the threat model}& \makecell[l]{$\bullet$ Deals with the cold-start problem.}\\ [1ex]
 \hline
 \end{tabular}}

\end{table*}

\subsection{Game Theory}
Game theory provides the framework to describe the strategic interaction between rational players. The nodes of a system can be seen as rational players since they try to maintain a high trust score to be selected as service providers. Therefore, some works model the trust management method as a game using game theoretic approaches. A summary of the results can be seen in Table \ref{tab:game}.

We can observe from the Table that all of the papers present experiments and define the Threat Model. Even though all the papers provide experiments, 25\% of them state the simulator used. Also, all of the papers provide a Time-Driven solution, but 25\% of them provide a hybrid scheme with both an Event and Time driven approach. None of the papers involve a centralized authority in their system. Regarding the Trust Formation: 25\% of them are using multiple trust parameters, 25\% only a single trust parameter, and 50\% do not specify the trust formation. 

Djedjig et al. \cite{Djedjig} proposed a distributed cooperation-trust-based
routing mechanism for RPL, where the malicious nodes can perform rank attacks and BA. At each hop of an RPL routing path, the child node selects the node that has a higher trust value, more energy, and better link quality as its preferred parent. The trust is calculated by taking into account energy consumption, honesty, selfishness, and the ETX.  Also, they translated the proposed trust management method into a strategy using game theory concepts. A non-trusted node will be discarded from the network. So, there is no advantage for a rational player to misbehave since it will be discarded from the network. The foundation of the solution is a non-zero-sum, non-cooperative iterated PD game. Experiments were performed using the Cooja simulator and were focused on comparing the proposed method with other schemes in terms of throughput, energy, Average Node Rank Changes under Blackhole and Rank attacks, and Average Packet Delivery Ratio. The results were presented as diagrams. The authors are focused on routing security and only deal with a set of trust-related attacks. They also test energy consumption, which is positive in a resource-constrained environment.

Esposito et al. \cite{ESPOSITO} proposed a system consisting of IoT nodes that communicate with edge nodes that participate in a blockchain. The blockchain stores a smart contract that periodically stores the trust scores. The smart contract receives the real numbers extracted from the nodes and uses fuzzy logic to translate the real numbers into linguistic terms.
The main threat to the system is to store on the blockchain a false value for the computed trust score. So, the solution is to focus on finding a way to reject these kinds of false messages. A game-theoretic approach was employed, forming a game between the edge node and a common node. A node has two available actions: to send a message or to avoid sending a message. This message might contain malicious data. At the reception of the message, the edge node can do only two
possible actions: Y indicating that it accepts the message and passes it to the blockchain participants to update its state, or N indicating that it rejects the message and does not pass it to the blockchain. Real sensors were used for performing the following experiments: comparison with other models, belief evolution, and attack success probability with and without the proposed defense. The final results of the experiments were presented in diagrams. The authors do not deal with blockchain-related attacks. Messages coming from untrusted sources are blocked.

Rani et al. \cite{Rani} considered several sensor nodes, deployed randomly
in a network field. All these nodes are equipped with limited-power batteries and have a short radio range. A base
station with an unlimited source of energy as a central administrative authority is also deployed in the network field. It is
also considered that the nodes of the network form clusters.
A cluster consists of cluster members and a cluster head. The proposed scheme uses evolutionary game theory in cluster formation and non-cooperative game theory to detect malicious
nodes in the network. When a node receives a trust request, it has two possible actions: to reply or not reply. When a node replies, it has some communication cost, which helps in energy efficiency. The malicious nodes can perform BMA, OOA, packet modification, collusion attacks, DoS, BA, and WHA. The experiments performed on the NS-3 simulator focused on: detection rate, average energy consumption, comparison with other schemes, trust evaluation time, and detection time. The results were presented in diagrams. The authors are presenting a cluster formation solution, which is a main issue in cluster-based networks. One drawback we identified is that the authors can expand their solution to cover more attacks.

Duan et al. \cite{Duan} adopted watchdog. Each sensor node is responsible for monitoring the behavior of its neighbors. A WSN was considered to consist of a few sink nodes and several sensor nodes. The main goal of this paper is to reduce energy consumption and latency for
trust evaluation. The paper proposed a method to find the optimal number of recommendations needed for trust evaluation while maintaining a high security level. The nodes are considered players with the following strategies: reply or not reply to save energy for trust computation. So this paper is proposing a dilemma game. The malicious nodes may perform BMA, DoS, or selfish attacks. The parameter used to calculate the trust is energy. NS-2 simulator was used to perform the following experiments: optimal selection of some values related to the trust process and comparison with other mechanisms. The results are presented in diagrams. One drawback of the solution is the overhead produced by the trust requests. On the other hand, the authors are examining energy consumption.

\begin{table*}[!t]\scriptsize
  \caption{Overview of approaches proposing Game Theoretic Trust Management methods}
  \label{tab:game}
  \centering
\resizebox{\textwidth}{!}{\begin{tabular}{||c | c | c | c | c | c | c | c | c | c||} 
 \hline
 Paper & \makecell{Info \\ Gathering}   & Trust Update & Experiments & Centralized & \makecell{Trust \\ Formation} & \makecell{Threat \\ Model} & Simulator & Limitations & Strengths  \\ [0.5ex] 
 \hline\hline
 \cite{Djedjig} & Both & Both & \checkmark &  & Multi-Trust & \checkmark& \checkmark& \makecell[l]{$\bullet$ Focus only on routing security. \\ $\bullet$ Deals only with a set of trust-related attacks} & \makecell[l]{$\bullet$ Different analysis for the game theoretic approach \\ $\bullet$ Tests energy consumption} \\ 
 \hline
 \cite{ESPOSITO} &  &  Time-Driven & \checkmark & & & \checkmark&  & \makecell[l]{$\bullet$ Do not study blockchain-related attacks}& \makecell[l]{$\bullet$ Blocks messages from untrusted nodes.}  \\
 \hline
 \cite{Rani} & Both & Time-Driven  & \checkmark & & &  \checkmark& \checkmark & \makecell[l]{$\bullet$ Expand the threat model} & \makecell[l]{$\bullet$ Cluster formation solution} \\
 \hline
  \cite{Duan} & Both & Time-Driven  & \checkmark & & Single-Trust &  \checkmark&  & \makecell[l]{$\bullet$ Overhead produced by trust request} & \makecell[l]{$\bullet$ Really focused on energy consumption} \\ [1ex]
 \hline
 \end{tabular}}

\end{table*}

\subsection{Context} 
One node can participate in many contexts and behave differently in each one. Some works take into account the different contexts to evaluate the trustworthiness of the node. The summary of the findings for this category is presented in Table \ref{tab:context}.

We can observe from the table that all of the papers use an Event-Driven approach, provide experiments, and use multiple parameters for trust calculation. 16.6\% of the papers use only direct observations for information gathering, while the rest use recommendations. Also, 16.6\% of the papers use a central authority for deriving trust. Even though all the papers provide experiments, only 33.34\% of them indicate the simulator used. Finally, 16.6\% of the papers define the threat model.

Rafey et al. \cite{Rafey} proposed a system where nodes form communities of interest. The proposed model takes social relationships into account to evaluate trust. Also, the trust is calculated in the different contexts in which the node is participating, and the final trust is the sum of the individual ones. This work also presents a way for storing trust values. The trust is derived from node transaction factors: Computational power, Context importance, Confidence, feedback, and social relationship factors: owner trust and SIoT relationship. The malicious entities can be: individual malevolent nodes, malevolent collectives, malevolent spies, malevolent pre-trusted nodes, partially malevolent collectives, or malevolent collectives with camouflage, and they can perform SA and BMA. The experiments were focused on performance and comparison with other mechanisms. The final results were presented in diagrams. The authors are dealing only with trust-related attacks. On the other hand, they propose a trust storage mechanism, which is helpful for resource-constrained devices. Moreover, they include social aspects in their solution.

Altaf et al. \cite{ALTAF} proposed a system consisting of users and service providers. The users are requesting services from the service providers. The context was used to calculate trust in a different context. One server has different trust scores for every context. Each edge node takes recommendations from context-similar nodes to calculate the trust of serving nodes. The trust parameters used are the following: server capability in terms of service provided, location, type of server, Quality of service, similarity with the recommender, location of the servers, and list of requested services. The experiments were focused on performance, resilience, and comparison with other models. The final results were presented in diagrams. The authors are filtering out the recommendations based on the context. Sometimes, a node might behave differently in different contexts. However, the authors do not specify the threat model.

Saied et al. \cite{SAIED} proposed a context-aware and multi-service trust management system. Upon a request from a node asking for assistance, the trust manager starts the entity selection process to return a set of trustworthy assisting nodes to the requester. A set of recommenders sends reports; the most important are those that lie to the same or more similar services and recent ones. A quality of recommendation score is assigned to each node, reflecting its trustworthiness when rating other nodes. The context was used to filter out recommendations and select the most relevant ones. The trust is calculated based on the following parameters:
The score is given by the requester node to the service provider evaluating the offered service, a weight that depends on time, and similarity, and quality of recommendations. Experiments were performed focused on the comparison of reactions against different kinds of attacks like on-off, bad-mouthing, and selective behavior attacks. The authors are filtering out the recommendations based on the context. Sometimes, a node might behave differently in different contexts. However, the authors do not specify the threat model, despite the fact that they are performing experiments on different attacks.

Abidi et al. \cite{Abidi} proposed a system consisting of nodes that create social relationships, a context Manager, a social relationship manager, and a trust formation adjustor. The goal of the system is to assist the nodes in finding trustworthy service providers. The level of trust between the service providers depends on both direct trust (the interactions between the requestor and the service provider) and, recommendations of the requestor's neighbors. The trust parameters adjust to the network context and the relationships between the nodes. The trust parameters taken into account are the quality of Service and social trust properties like honesty, cooperativeness, and social relationships. Trust calculations rely on Social relationship factors.  The following are the social relationships that are formed in the model: Parental Object Relationship, co-location Object Relationship, co-work Object Relationship, ownership Object Relationship, and social object relationships. The experiments were performed in MATLAB and they were focused on the performance of the method and comparison with other models. The final results were presented in diagrams. The authors do not specify the threat model, but they take social aspects into account.

Adewuyi et al. \cite{Adewuyi} proposed a system called CTRUST. In this work, the authors model the trust units with mathematical functions. The trust properties used to calculate the trust level are the social relationships between the nodes and the context. This paper also introduces a parameter to model trust maturity, the point at which trust can be computed using direct interactions alone. The performance was evaluated based on trust accuracy, convergence, and resiliency. Diagrams present the final results. The authors do not specify the threat model, but they take social aspects into account.

Magdich et al. \cite{Magdich1} are considering an environment consisting of a set of users and a set of devices. A user can own one or multiple devices, and the devices can provide or request one or more services. Every device is represented by a vector containing three values: user, device, and environment (public or private). The environment value sets the threshold of trust. Also, each device stores its characteristics (manufacturer, type, capacity, and location), the profile of its owners  (friendship, CoI, and Co-work), the transaction history between other nodes, and trust values. In this approach, the trust of the owner, the device, and the environment are taken into account to decide the trust value. The method also uses recommendations. Afterward, the threshold of trust has to be decided. The authors are proposing a Machine Learning technique that they compare with the static method, available in the literature. The ML (Artificial Neural Network) algorithm classifies the nodes as trustworthy or not. After each interaction, the nodes evaluate each other and share the result with the other nodes as a recommendation. The authors do not specify the threat model, but they take social aspects into account.

\begin{table*}[!t]\scriptsize
  \caption{Overview of approaches proposing Context based Trust Management methods}
  \label{tab:context}
  \centering
\resizebox{\textwidth}{!}{\begin{tabular}{||c | c | c | c | c | c | c | c | c | c||} 
 \hline
 Paper & \makecell{Info \\ Gathering}   & Trust Update & Experiments & Centralized & \makecell{Trust \\ Formation} & \makecell{Threat \\ Model} & Simulator & Limitations & Strengths  \\ [0.5ex] 
 \hline\hline
 \cite{Rafey} & Both & Event-Driven & \checkmark &  & Multi-Trust & \checkmark & & \makecell[l]{$\bullet$ Only trust-related attacks}& \makecell[l]{$\bullet$ Proposal for trust storage \\ $\bullet$ Takes into account social aspects} \\ 
 \hline
 \cite{ALTAF} & Both & Event-Driven & \checkmark &  & Multi-Trust & & & \makecell[l]{$\bullet$ Do not specify the threat model} & \makecell[l]{$\bullet$ Filters out dissimilar recommendations} \\
 \hline
 \cite{SAIED} & Direct   &  Event-Driven & \checkmark & \checkmark& Multi-Trust &  &  & \makecell[l]{$\bullet$ Do not specify the threat model} & \makecell[l]{$\bullet$ Filters out dissimilar recommendations} \\
 \hline
  \cite{Abidi} & Both & Event-Driven & \checkmark &  & Multi-Trust & & \checkmark & \makecell[l]{$\bullet$ Do not specify the threat model} & \makecell[l]{$\bullet$ Takes into account social aspects}\\ 
 \hline
   \cite{Adewuyi} & Both & Event-Driven & \checkmark &  & Multi-Trust & &  & \makecell[l]{$\bullet$ Do not specify the threat model} & \makecell[l]{$\bullet$ Takes into account social aspects} \\
   \hline
   \cite{Magdich1} & Both & Event-Driven & \checkmark &  & Multi-Trust & & \checkmark & \makecell[l]{$\bullet$ Do not specify the threat model} & \makecell[l]{$\bullet$ Takes into account social aspects}\\ [1ex]
 \hline
 \end{tabular}}

\end{table*}

\subsection{Social}
Social IoT provides a combination of IoT and social networking. The sensors can establish social relationships \cite{Roopa}. Some works exploit this aspect to create trust management methods. A summary of the findings for this category can be seen in Table \ref{tab:social}.

We can observe from the table that 95.6\% of the papers use both direct and indirect trust, and only 4.4\% use only direct observations. Regarding the Trust Update 65.2\% of the papers use the Event-Driven approach, 13\% use the Time-Driven approach, and 13\% do not refer to the Trust Update. It is worth pointing out that 4.4\% of the papers use a hybrid approach to the Trust Update. Also, 95.6\% are presenting Experiments but only 30.4\% of them are referring to the simulator used. 8.7\% of the papers preferred a Centralized authority to manage the trust calculation procedure. Regarding Trust Formation: 8.7\% of the papers are using the Single-Trust approach, while the rest are using the Multi-Trust approach. Finally, 52.2\% of the papers define the Threat Model.

Rafey et al. \cite{Rafey} proposed a system where nodes form communities of interest. The proposed model takes social relationships into account to evaluate trust. Also, the trust is calculated in the different contexts in which the node is participating, and the final trust is the sum of the individual ones. This work also presents a way for storing trust values. The trust is derived from node transaction factors: Computational power, Context importance, Confidence, feedback, and social relationship factors: owner trust and SIoT relationship. The malicious entities can be: individual malevolent nodes, malevolent collectives, malevolent spies, malevolent pre-trusted nodes, partially malevolent collectives, or malevolent collectives with camouflage, and they can perform SA and BMA. The experiments were focused on performance and comparison with other mechanisms. The final results were presented in diagrams. The authors are dealing only with trust-related attacks. On the other hand, they propose a trust storage mechanism, which is helpful for resource-constrained devices. Moreover, they include social aspects in their solution.

Nitti et al. \cite{Nitti} proposed a system consisting of a network of nodes, several pre-trusted entities to hold a distributed hash table structure, and four other components to manage the network: relationship management, service discovery, service composition, and trustworthiness management. This paper defined two models for trustworthiness management: subjective and objective. In the first model, each node computes the trustworthiness of its friends using its own experience and the opinions of the friends in common. In the second model, the information about each node is stored on a distributed hash table structure. The following trust parameters are taken into account: feedback, the total number of transactions, the credibility, the transaction factor, the relationship factor, the notion of centrality, and computation capability. Trust calculations rely on Social relationship factors.  The following are the social relationships that are formed in the model: Parental Object Relationship, co-location Object Relationship, co-work Object Relationship, ownership Object Relationship, and social object relationship. The experiments conducted focus on comparing the performance with other models and how the proposed approaches work with three different dynamic behaviors of the nodes. This approach requires some pre-trusted entities to be involved, which is not always applicable in real life. Also, they do not provide the threat model. Finally, a pro is that they provide a view of trust of the whole system as a holistic evaluation.

Mon et al. \cite{Mon} proposed a cluster-based system with a central trust entity. Initially, a cluster is formed, and the master node is selected based on its trust scores, which include QoS and Social trust properties. The master node is periodically updated based on the trust value using regression model-based clustering. Experiments were conducted to test the performance of the proposed approach. The results were presented in diagrams. One limitation is that trust values are required to form clusters; it is not clear what happens during bootstrapping. One strength of the solution is that it takes into account data trust, which is important for identifying false messages even from trustworthy nodes.

Marche et al. \cite{Marche} proposed a system that focuses on detecting trust attacks. To achieve it, machine learning techniques are applied. Trust calculations rely on Social relationship factors.  The following are the social relationships that are formed in the model: Parental Object Relationship, co-location Object Relationship, co-work Object Relationship, ownership Object Relationship, and social object relationship. A malicious node can be malicious to everyone or selectively and perform the following attacks: OOA, WHA, BMA, BSA, SA, and OSA. For trust computations, the following parameters are taken into account: previous interactions, computation capabilities, relationship factors, external opinions, and dynamic knowledge. Experiments were conducted to test the performance of the iSVM. Diagrams present the results of the experiments. One limitation is that the approach only deals with trust-related attacks. On the other hand, at the steady state, only one parameter is required to compute the trust value, which saves time and energy.

Abidi et al. \cite{Abidi} proposed a system consisting of nodes that create social relationships, a context Manager, a social relationship manager, and a trust formation adjustor. The goal of the system is to assist the nodes in finding trustworthy service providers. The level of trust between the service providers depends on both direct trust (the interactions between the requestor and the service provider) and, recommendations of the requestor's neighbors. The trust parameters adjust to the network context and the relationships between the nodes. The trust parameters taken into account are the quality of Service and social trust properties like honesty, cooperativeness, and social relationships. Trust calculations rely on Social relationship factors.  The following are the social relationships that are formed in the model: Parental Object Relationship, co-location Object Relationship, co-work Object Relationship, ownership Object Relationship, and social object relationships. The experiments were performed in MATLAB and they were focused on the performance of the method and comparison with other models. The final results were presented in diagrams. The authors do not specify the threat model, but they take into account the context which is a strength since nodes might act differently in different contexts.

Nitti et al. \cite{NittiSUB} proposed a system where the nodes evaluate the trustworthiness of other nodes based on their observations and recommendations of common friends (between the trustor and the trustee). The trust parameters taken into account are feedback, the total number of transactions, the credibility, the transaction factor, the relationship factor, the notion of centrality, and computation capability. Trust calculations rely on Social relationship factors.  The following are the social relationships that are formed in the model: Parental Object Relationship, co-location Object Relationship, co-work Object Relationship, ownership Object Relationship, and social object relationship. Experiments were conducted to test the performance of the proposed solution. The results are presented in diagrams. The limitation of this solution is that the authors do not specify the threat model.

Wang et al. \cite{Wang} proposed a trust model based on direct and indirect trust computation with trust prediction. The prediction method depends on the combination of exponential smoothing and a Markov chain. Exponential smoothing was employed to predict trust and a Markov chain was employed to fix any deviation. Thus, a prediction method was employed to predict the current trust level based on interaction history, behavior history, and some other factors like the device model. For the trust computation, both social and unsocial parameters were considered. The following experiments were performed: comparison of trust prediction with different exponential smoothing coefficients; comparison between first and second exponential smoothing; and experiments for different kinds of attacks. The results were presented as diagrams. The authors do not specify the threat model, which is a drawback. On the other hand, they are proposing a solution to the communication latency issue.

Chen et al. \cite{ChenSOA} designed and analyzed a trust management protocol for SOA-based IoT systems. The IoT owners can share their feedback, so a filtering method was proposed to select the feedback of owners with common interests. Also, the nodes can adjust the weights of direct trust and recommendations. The social aspects were used to weigh the recommendations: Friendship, social contact, and community of interest. User satisfaction is the trust parameter used to calculate trust. They also introduced a method for trust storage. The malicious nodes can perform BMA, SPA, BSA, and OSA. The experiments were conducted on the NS-3 simulator, and they tested convergence, accuracy, resiliency, the effectiveness of storage management protcols, and comparative analysis. The results are presented in diagrams. The main limitation of this approach is the threat model, which only includes trust-related attacks. On the other hand, the authors are proposing a trust-based storage solution for resource-constrained devices.

Bao et al. \cite{Bao} proposed a dynamic trust management protocol for IoT systems. The trust evaluation takes into account direct and indirect recommendations and social aspects. They also take into account scalability in terms of storing trust values. The malicious nodes can perform BMA, SPA, and BSA. The nodes form communities, and during the trust evaluation procedure, the social aspects considered are honesty, cooperativeness, and the community of interest. Experiments were conducted to observe the effect of some weights used in the trust evaluation and the protocol's resiliency to trust attacks. The results are presented in diagrams. One limitation of this approach is the threat model, which only includes trust-related attacks. On the other hand, the authors are testing the scalability of storing trust values.

Kowshalya et al. \cite{Kowshalya} presented a system where the network is presented as a graph. Where V are the participants and E represents the edges between them. The devices form communities of interest based on parental, co-work, and co-location relationships. Also, this paper ensures secure communication among SIoT nodes through simple secret codes. For the trust evaluation procedure, the properties taken into account are the following: honesty, cooperativeness, community of interest, and energy. For the experiments, the SWIM platform was used, and the experiments tested the performance of the proposed model and compared it with other schemes. The results were presented in diagrams. During the experiments, the effect of the trust weights was analyzed. The authors do not specify the threat model, which constitutes a drawback. But on the other hand, they are proposing a way to ensure secure communications using secret codes.

Adewuyi et al. \cite{Adewuyi} proposed a system called CTRUST. In this work, the authors model the trust units with mathematical functions. The trust properties used to calculate the trust level are the social relationships between the nodes and the context. This paper also introduces a parameter to model trust maturity, the point at which trust can be computed using direct interactions alone. The performance was evaluated based on trust accuracy, convergence, and resiliency. Diagrams present the final results. The authors do not specify the threat model, but they take into account the context.

Bao et al. \cite{BaoSCA} proposed a system where the nodes form communities of interest. The protocol is distributed and each node evaluates the trust of nodes that share interests. The system can adapt to changes in communities of interest by dynamically selecting the trust parameters. For scalability, the authors also proposed a storage management strategy to save memory from the resource constraint of IoT devices. The malicious nodes can perform BMA, SPA, and BSA. The experiments tested the effect of changing some weight values related to the trust evaluation procedure and the trust evaluation with limited storage space. The results are presented in diagrams. The main limitation of this approach is the threat model, which only includes trust-related attacks. On the other hand, the authors are proposing a trust-based storage solution for resource-constrained devices.

Das et al. \cite{Das} proposed a system consisting of IoT nodes and Fog nodes. The IoT nodes communicate with the closest Fog node. This paper proposes a community-based trust management architecture by considering self-trust, social trust, green trust, and QoS trust. Experiments were conducted on MATLAB concerning the performance of the system. Diagrams present the final results of the experiments. The authors do not specify the threat model. On the other side, they are testing the energy consumption of their solution.

Awan et al. \cite{Awan} proposed a multilevel architecture system. The nodes form communities of interest. Every community has a server to calculate trust. A set of communities forms a domain that has a server to calculate the trust of the domain. The whole system is governed by a server that is responsible for the trust of all the domains. The trust properties taken into account are compatibility, honesty, and competence. The authors are presenting a purely theoretical model in which they do not specify the threat model. On the other side, they are proposing cross-domain trust management, taking into account different domains. 

Amiri-Zarandi et al. \cite{Amiri-Zarandi} proposed a fog architecture scheme. The interactions between the nodes and the blockchain can occur directly or via edge nodes. The nodes are divided into clusters, and they communicate with a fog device. The social connection between the devices is used for trust evaluation. The scheme also works with recommendations that are filtered by a lightweight algorithm. The blockchain is used for storing trust-related data. The malicious nodes can perform BMA, BSA, SFA, and DoS. Honesty was used as the trust parameter. Also, the following experiments took place using the Ethereum blockchain: performance evaluation and experimental comparison with other models. The results were presented in diagrams and tables. The authors do not study any blockchain-related attacks. This approach dynamically selects counselors, as one node might be a good fit at first but behave maliciously afterward. Also, they perform cost analysis in terms of gas.

Jayasinghe et al. \cite{Jayasinghe} proposed a system where the nodes form communities of interest. In this model, the transactions are under evaluation, and should be determined if a transaction is trustworthy. The trust parameters used are co-location relationships, co-work relationships, mutuality and centrality, and cooperativeness. An unsupervised learning technique was employed to label the data's trustworthiness. After the labeling, an SVM model predicts the trust level. Experiments were performed to observe the performance of the proposed solution. The authors do not specify the threat model. They offer a metric that provides a perception of a node before interaction. This can be used when there have been no previous interactions or a new node has just entered the network.

Bao et al. \cite{BaoAPP} consider an IoT environment with no centralized trusted authority. Every device (node) has an owner, and an owner could have many devices. Each owner has a list of friends, representing their social relationships. The trust properties used for trust calculation are honesty, cooperativeness, and community interest. The threat model of this approach includes BMA, SPA, and BSA. The experiments tested the effect of trust parameters on trust evaluation. One drawback of this approach is the limited threat model to only trust-related attacks.

Chen et al. \cite{Chen_trust-basedservice} proposed a trust management protocol for Social IoT systems that can form a community of interests. The trust parameters can be dynamically adapted to changes in the environment. The malicious nodes can perform discrimination attacks: BMA, SPA, WHA, and BSA. Each device has an owner. Each owner has a list of friends, representing their social relationships. The trust parameters taken into account for the trust evaluation procedure are honesty, cooperativeness, and community of interest. Experiments were conducted on the NS-3 simulator to test the performance of the proposed protocol. The main limitation of the approach is that the threat model is limited to a set of trust-related attacks.

Wen et al. \cite{Wen} proposed a method based on \cite{Abderrahim}, a cluster-based scheme, where the nodes form communities of interest. The whole network is governed by a SIOT server. In this work, trust is evaluated from both direct and indirect trust. They introduced a deep learning model to predict the trust value of the new nodes to solve the cold-start problem. The malicious nodes can perform BMA, BSA, and OOA. The following experiments took place: accuracy with a different number of malicious nodes, comparison with another model, and adding a new node to the network. The final results were presented in diagrams. The main limitation of the approach is that the threat model is limited to a set of trust-related attacks. On the other side, the approach deals with the cold-start problem, an important issue for newcomers in a network.

Abderrahim et al. \cite{Abderrahim} proposed a cluster approach. The nodes form communities of interest. The network is governed by a SIOT server. This approach detects OOA through the use of a Kalman Filter. The malicious nodes can perform BMA, BSA, and OOA. The outcome of the transactions is taken into account to calculate the trust, as are the previous trust values. The code was developed using Python programming language to identify the best weights under which the estimated trust is close to the objective one, observe the performance during OOA, and perform experiments on trust prediction. The results are presented in diagrams. The main limitation of the approach is that the threat model is limited to a set of trust-related attacks.

Aalibagi et al. \cite{Aalibagi} presented a social network using a bipartite graph. Assuming that there are a finite number of service types, they construct one for every service. The bipartite graph consists of two sets of nodes: trustors $U$ and trustees $V$. In the bipartite graph, trustor $u$ has an edge linked to trustee $v$, if $u$ has already used the services provided by $v$ at least once. The edge is decorated with a weight (number) indicating the trustor’s trust experience while using services provided by the trustee. As a next step, they are finding trust similarities between trustors based on their past experiences, similarity, and centrality measures. To measure the similarity, the Hellinger similarity, the Bayesian similarity, and the connection similarity are used. Also, two other metrics for centrality are used. Afterward, matrix factorization is used to predict the trustworthiness of a trustee. The method takes into account the friend’s feedback based on how similar the two nodes are. This method also considers the cold-start problem.  The main limitation of the approach is that the threat model is limited to a set of trust-related attacks. On the other side, the approach deals with the cold-start problem, an important issue for newcomers in a network.

Magdich et al. \cite{MAGDICH202292} are working on a network of service requestors (SR) and service providers (SP), where the SR evaluates the trust experience with the SP. The trust computation relies on QoS and social metrics. The nodes are also taking into account recommendations from other nodes. After forming the trust score, the nodes act for attack detection. The goal is to identify the attacker and predict the attack it's performing. To achieve it Machine learning and Deep Learning techniques were applied using the following features: reputation, recommendation, similarity, knowledge, and trust. Experiments were performed using the Cooja simulator. The main limitation of the approach is that the threat model is limited to a set of trust-related attacks. On the other side, the approach is tested on real social data and experiments made with different Machine learning and Deep Learning techniques.

Magdich et al. \cite{Magdich1} are considering an environment consisting of a set of users and a set of devices. A user can own one or multiple devices, and the devices can provide or request one or more services. Every device is represented by a vector containing three values: user, device, and environment (public or private). The environment value sets the threshold of trust. Also, each device stores its characteristics (manufacturer, type, capacity, and location), the profile of its owners  (friendship, CoI, and Co-work), the transaction history between other nodes, and trust values. In this approach, the trust of the owner, the device, and the environment are taken into account to decide the trust value. The method also uses recommendations. Afterward, the threshold of trust has to be decided. The authors are proposing a Machine Learning technique that they compare with the static method available in the literature. The ML (Artificial Neural Network) algorithm classifies the nodes as trustworthy or not. After each interaction, the nodes evaluate each other and share the result with the other nodes as a recommendation. The authors do not specify the threat model, but they take the context into account.

\begin{table*}[!t]\scriptsize
  \caption{Overview of approaches proposing Social based Trust Management methods}
  \label{tab:social}
  \centering
\resizebox{\textwidth}{!}{\begin{tabular}{||c | c | c | c | c | c | c | c | c | c||} 
 \hline
 Paper & \makecell{Info \\ Gathering}   & Trust Update & Experiments & Centralized & \makecell{Trust \\ Formation} & \makecell{Threat \\ Model} & Simulator & Limitations & Strengths  \\ [0.5ex] 
 \hline\hline
  \cite{Rafey} & Both & Event-Driven & \checkmark &  & Multi-Trust & \checkmark & & \makecell[l]{$\bullet$ Only trust-related attacks}& \makecell[l]{$\bullet$ Proposal for trust storage \\ $\bullet$ Takes into account the context}  \\ 
 \hline
 \cite{Nitti} & Both &  Event-Driven & \checkmark& & Multi-Trust & & &\makecell[l]{$\bullet$ Pretrusted entities are involved \\ $\bullet$ Do not provide the threat model} &\makecell[l]{$\bullet$ Provides a view of the trust of the whole system }  \\
 \hline
 \cite{Mon} & Both &  Time-Driven & \checkmark  & \checkmark& Multi-Trust &  &  & \makecell[l]{$\bullet$ Bootstrapping issue with forming clusters \\ the trust values are required \\ $\bullet$ Do not provide the threat model} & \makecell[l]{$\bullet$ Takes into account the data trust}\\
 \hline
  \cite{Marche} & Both &  Event-Driven & \checkmark& & Multi-Trust & \checkmark & & \makecell[l]{$\bullet$ Deals only with trust-related attacks}& \makecell[l]{$\bullet$ At steady-state only one \\ parameter is used to compute trust} \\
 \hline
  \cite{Abidi} & Both & Event-Driven & \checkmark &  & Multi-Trust & & \checkmark & \makecell[l]{$\bullet$ Do not specify the threat model} & \makecell[l]{$\bullet$ Takes into account the context}\\
  \hline
    \cite{NittiSUB} & Both &  Event-Driven & \checkmark& & Multi-Trust & & &\makecell[l]{$\bullet$ Do not specify the threat model} & \\
  \hline
   \cite{Wang} & Both &   & \checkmark& & Multi-Trust & & & \makecell[l]{$\bullet$ Do not specify the threat model}&\makecell[l]{$\bullet$ Dealing with the communication latency} \\ 
 \hline
 \cite{ChenSOA} & Both &  Event-Driven & \checkmark& & Single-Trust &\checkmark & \checkmark& \makecell[l]{$\bullet$ Deals only with a set of trust-related attacks}& \makecell[l]{$\bullet$ Section for trust storage} \\ 
 \hline
  \cite{Bao} & Both &  Event-Driven & \checkmark& & Multi-Trust & \checkmark & & \makecell[l]{$\bullet$ Deals only with a set of trust-related attacks} & \makecell[l]{$\bullet$ Tests scalability for storing the trust values}\\ 
 \hline
   \cite{Kowshalya} & Both &  Event-Driven & \checkmark& & Multi-Trust &  & &\makecell[l]{$\bullet$ Do not specify the threat model} & \makecell[c]{$\bullet$ Ensure secure communication using secret codes}\\ 
 \hline
   \cite{Adewuyi} & Both & Event-Driven & \checkmark &  & Multi-Trust & &  & \makecell[l]{$\bullet$ Do not specify the threat model} & \makecell[l]{$\bullet$ Takes into account the context} \\ 
 \hline
   \cite{BaoSCA} & Both &  Event-Driven & \checkmark& &  & \checkmark & & \makecell[l]{$\bullet$ Deals only with a set of trust-related attacks} & \makecell[l]{$\bullet$ Propose a storage management strategy}\\ 
   \hline
   \cite{Das} & Both &  & \checkmark& & Multi-Trust &  &\checkmark & \makecell[l]{$\bullet$ Do not specify the threat model} & \makecell[l]{$\bullet$ Tests the energy consumption}\\ 
   \hline
   \cite{Awan} & Both & Both & & \checkmark & Multi-Trust &  & & \makecell[l]{$\bullet$ Purely theoretical model \\ $\bullet$ Do not specify the threat model}& \makecell[l]{$\bullet$ Cross-domain trust management \\ $\bullet$ Extensive literature review}\\ 
   \hline
  \cite{Amiri-Zarandi} & Both & Event-Driven & \checkmark& & Single-Trust & \checkmark & & \makecell[l]{$\bullet$ Do not study blockchain-related attacks} & \makecell[l]{$\bullet$ Dynamically selected counselors \\ $\bullet$ Cost analysis in terms of gas}\\ 
  \hline
    \cite{Jayasinghe} & Direct &  &\checkmark & & Multi-Trust &  & & \makecell[l]{$\bullet$ Do not specify the threat model} & \makecell[l]{$\bullet$ Metric that provides a perception \\ of a node before interaction} \\
 \hline
 \cite{BaoAPP} & Both & Event-Driven &\checkmark & & Multi-Trust & \checkmark  & & \makecell[l]{$\bullet$ Deals only with a set of trust-related attacks} & \\
 \hline
  \cite{Chen_trust-basedservice} & Both & Event-Driven &\checkmark & & Multi-Trust & \checkmark  & \checkmark & \makecell[l]{$\bullet$ Deals only with a set of trust-related attacks} & \makecell[l]{$\bullet$ Dynamically adjustable} \\
  \hline
   \cite{Wen} & Both & Event-Driven  & \checkmark& & Multi-Trust&  \checkmark &  & \makecell[l]{$\bullet$ Deals only with a set of trust-related attacks} & \makecell[l]{$\bullet$ Deals with cold-start problem } \\
 \hline
    \cite{Abderrahim} & Both &  Time-Driven & \checkmark& & Multi-Trust&  \checkmark & \checkmark & \makecell[l]{$\bullet$ Deals only with a set of trust-related attacks} & \makecell[l]{$\bullet$ }\\
    \hline
  \cite{Aalibagi} & Both &  &\checkmark & & Multi-Trust &  \checkmark &  & \makecell[l]{$\bullet$ Deals only with a set of trust-related attacks} & \makecell[l]{$\bullet$ Deals with cold-start problem}\\ 
  \hline
  \cite{MAGDICH202292} & Both & Time-Driven &\checkmark & & Multi-Trust &  \checkmark & \checkmark & \makecell[l]{$\bullet$ Deals only with a set of trust-related attacks} & \makecell[l]{$\bullet$ Experiments with real social data \\ $\bullet$ Different Models were tested}\\
   \hline
   \cite{Magdich1} & Both & Event-Driven & \checkmark &  & Multi-Trust & & \checkmark & \makecell[l]{$\bullet$ Do not specify the threat model} & \makecell[l]{$\bullet$ Takes into account the context}\\ [1ex]
    \hline
 \end{tabular}}

\end{table*}

\subsection{Prediction} 
Sometimes the image of the trustworthiness of a node is not clear, so a prediction mechanism would help the rest of the nodes estimate its trust level. There are works for trust management that use prediction mechanisms to enhance the trust evaluation process. The summary of the findings for this category is presented in Table \ref{tab:prediction}.

We can observe from the Table that 22.2\% of the paper uses only direct trust for information gathering, while the rest also uses recommendations.  Regarding the Trust Update: 22.2\% of the papers use a Time-Driven approach, 33.3\% use an Event-Driven approach, and the rest do not refer to the Trust Update. All of the papers present Experiments, but 55.5\% of them refer to the simulator used. Also, all of the papers use Multi-Trust for Trust Formation, and none of them use a Central Authority for trust calculation. Finally, 55.5\% of the papers define the Threat Model.

Wang et al. \cite{Wang} proposed a trust model based on direct and indirect trust computation with trust prediction. The prediction method depends on the combination of exponential smoothing and a Markov chain. Exponential smoothing was employed to predict trust and a Markov chain was employed to fix any deviation. Thus, a prediction method was employed to predict the current trust level based on interaction history, behavior history, and some other factors like the device model. For the trust computation, both social and unsocial parameters were considered. The following experiments were performed: comparison of trust prediction with different exponential smoothing coefficients; comparison between first and second exponential smoothing; and experiments for different kinds of attacks. The results were presented as diagrams. The authors do not specify the threat model, which is a drawback. On the other hand, they are proposing a solution to the communication latency issue.

Subhash et al. \cite{Subhash} proposed the Power Trust. Power Trust assigns trust values to the nodes of the network based on energy auditing. Using the energy auditing model, they calculate the trust values of every node present in the network dynamically and predict physical and cyberattacks. To detect the attacks, a deep learning model was trained with past data that contains normal and excessive energy consumption due to an attack. The model can predict both physical and cyberattacks. The experiments were performed in the Cooja simulator, and they were focused on the performance and the accuracy of the method. The results were presented in diagrams. The authors do not state the threat model, and they do not give details about the deep learning model used. The method takes into account energy consumption, which is important in a resource-constrained environment. Finally, they also predict both physical and cyberattacks.

Wen et al. \cite{Wen} proposed a method based on \cite{Abderrahim}, a cluster-based scheme, where the nodes form communities of interest. The whole network is governed by a SIOT server. In this work, trust is evaluated from both direct and indirect trust. They introduced a deep learning model to predict the trust value of the new nodes to solve the cold-start problem. The malicious nodes can perform BMA, BSA, and OOA. The following experiments took place: accuracy with a different number of malicious nodes, comparison with another model, and adding a new node to the network. The final results were presented in diagrams. The main limitation of the approach is that the threat model is limited to a set of trust-related attacks. On the other side, the approach deals with the cold-start problem, an important issue for newcomers in a network.

Alnumay et al. \cite{Alnumay} proposed a cluster-based system where every cluster has a cluster head. The proposed trust model combines both direct and indirect trust. A Beta probabilistic distribution is used, and the theory of ARMA/GARCH to combine the trust units and derive the trust value. The cluster heads can predict the trust value ahead using this method. A malicious node can perform the following attacks: SFA, Routing table overflow and resource consumption attacks, Byzantine, BA, DoS attacks and SDA. The trust is computed based on the number of packets properly forwarded, the number of packets dropped, and the number of packets falsely injected. Experiments were performed to observe the performance and accuracy of the proposed solution. Also, they presented some diagrams depicting the comparison with two other models. The main limitation of the approach is that during the initialization of the clusters, a malicious node might become cluster head at first. On the other side, the authors are dealing with a high-mobility environment and can predict multiple steps ahead.

Abderrahim et al. \cite{Abderrahim} proposed a cluster approach. The nodes form communities of interest. The network is governed by a SIOT server. This approach detects OOA through the use of a Kalman Filter. The malicious nodes can perform BMA, BSA, and OOA. The outcome of the transactions is taken into account to calculate the trust, as are the previous trust values. The code was developed using Python programming language to identify the best weights under which the estimated trust is close to the objective one, observe the performance during OOA, and perform experiments on trust prediction. The results are presented in diagrams. The main limitation of the approach is that the threat model is limited to a set of trust-related attacks.

Jayasinghe et al. \cite{Jayasinghe} proposed a system where the nodes form communities of interest. In this model, the transactions are under evaluation, and should be determined if a transaction is trustworthy. The trust parameters used are co-location relationships, co-work relationships, mutuality and centrality, and cooperativeness. An unsupervised learning technique was employed to label the data's trustworthiness. After the labeling, an SVM model predicts the trust level. Experiments were performed to observe the performance of the proposed solution. The authors do not specify the threat model. They offer a metric that provides a perception of a node before interaction. This can be used when there have been no previous interactions or a new node has just entered the network.

Aalibagi et al. \cite{Aalibagi} presented a social network using a bipartite graph. Assuming that there are a finite number of service types, they construct one for every service. The bipartite graph consists of two sets of nodes: trustors $U$ and trustees $V$. In the bipartite graph, trustor $u$ has an edge linked to trustee $v$, if $u$ has already used the services provided by $v$ at least once. The edge is decorated with a weight (number) indicating the trustor’s trust experience while using services provided by the trustee. As a next step, they are finding trust similarities between trustors based on their past experiences, similarity, and centrality measures. To measure the similarity, the Hellinger similarity, the Bayesian similarity, and the connection similarity are used. Also, two other metrics for centrality are used. Afterward, matrix factorization is used to predict the trustworthiness of a trustee. The method takes into account the friend’s feedback based on how similar the two nodes are. This method also considers the cold-start problem.  The main limitation of the approach is that the threat model is limited to a set of trust-related attacks. On the other side, the approach deals with the cold-start problem, an important issue for newcomers in a network.

Magdich et al. \cite{MAGDICH202292} are working on a network of service requestors (SR) and service providers (SP), where the SR evaluates the trust experience with the SP. The trust computation relies on QoS and social metrics. The nodes are also taking into account recommendations from other nodes. After forming the trust score, the nodes act for attack detection. The goal is to identify the attacker and predict the attack it's performing. To achieve it Machine learning and Deep Learning techniques were applied using the following features: reputation, recommendation, similarity, knowledge, and trust. Experiments were performed using the Cooja simulator. The main limitation of the approach is that the threat model is limited to a set of trust-related attacks. On the other side, the approach is tested on real social data and experiments made with different Machine learning and Deep Learning techniques.

Magdich et al. \cite{Magdich1} are considering an environment consisting of a set of users and a set of devices. A user can own one or multiple devices, and the devices can provide or request one or more services. Every device is represented by a vector containing three values: user, device, and environment (public or private). The environment value sets the threshold of trust. Also, each device stores its characteristics (manufacturer, type, capacity, and location), the profile of its owners  (friendship, CoI, and Co-work), the transaction history between other nodes, and trust values. In this approach, the trust of the owner, the device, and the environment are taken into account to decide the trust value. The method also uses recommendations. Afterward, the threshold of trust has to be decided. The authors are proposing a Machine Learning technique that they compare with the static method available in the literature. The ML (Artificial Neural Network) algorithm classifies the nodes as trustworthy or not. After each interaction, the nodes evaluate each other and share the result with the other nodes as a recommendation. The authors do not specify the threat model, but they take the context and social aspects into account.

\begin{table*}[!t] \scriptsize
  \caption{Overview of approaches proposing Prediction based Trust Management methods}
  \label{tab:prediction}
  \centering
\resizebox{\textwidth}{!}{\begin{tabular}{||c | c | c | c | c | c | c | c | c | c||} 
 \hline
 Paper & \makecell{Info \\ Gathering}   & Trust Update & Experiments & Centralized & \makecell{Trust \\ Formation} & \makecell{Threat \\ Model} & Simulator & Limitations & Strengths  \\ [0.5ex] 
 \hline\hline
 \cite{Wang} & Both &   & \checkmark& & Multi-Trust & & & \makecell[l]{$\bullet$ Do not specify the threat model}&\makecell[l]{$\bullet$ Dealing with the communication latency}   \\ 
 \hline
 \cite{Subhash} & Direct  &  & \checkmark& & Multi-Trust &  & \checkmark & \makecell[l]{$\bullet$ Do not specify the threat model \\ $\bullet$ Do not give details about the deep learning}& \makecell[l]{$\bullet$ Predict the physical attacks and cyber-attacks \\ $\bullet$ Focused on energy consumption}\\
 \hline
 \cite{Wen} & Both & Event-Driven  & \checkmark& & Multi-Trust&  \checkmark &  & \makecell[l]{$\bullet$ Deals only with a set of trust-related attacks} & \makecell[l]{$\bullet$ Deals with cold-start problem \\ $\bullet$ Takes into account social aspects} \\
 \hline
  \cite{Alnumay} & Both &  Event-Driven & \checkmark& & Multi-Trust&  \checkmark &  & \makecell[l]{$\bullet$At initialization a malicious node can be CH} & \makecell[l]{$\bullet$ Predicts multi-step ahead \\ $\bullet$ Mobility} \\
 \hline
    \cite{Abderrahim} & Both &  Time-Driven & \checkmark& & Multi-Trust&  \checkmark & \checkmark & \makecell[l]{$\bullet$ Deals only with a set of trust-related attacks} & \makecell[l]{$\bullet$ Takes into account social aspects}\\ 
 \hline
  \cite{Jayasinghe} & Direct &  &\checkmark & & Multi-Trust &  & & \makecell[l]{$\bullet$ Do not specify the threat model} & \makecell[l]{$\bullet$ Metric that provides a perception \\ of a node before interaction}\\ \hline
  \cite{Aalibagi} & Both &  &\checkmark & & Multi-Trust &  \checkmark & \checkmark & \makecell[l]{$\bullet$ Deals only with a set of trust-related attacks} & \makecell[l]{$\bullet$ Deals with cold-start problem}\\ 
  \hline
    \cite{MAGDICH202292} & Both & Time-Driven &\checkmark & & Multi-Trust &  \checkmark & \checkmark & \makecell[l]{$\bullet$ Deals only with a set of trust-related attacks} & \makecell[l]{$\bullet$ Experiments with real social data \\ $\bullet$ Different Models were tested}\\
   \hline
   \cite{Magdich1} & Both & Event-Driven & \checkmark &  & Multi-Trust & & \checkmark & \makecell[l]{$\bullet$ Do not specify the threat model} & \makecell[l]{$\bullet$ Takes into account social aspects and the context}\\ [1ex]
 \hline
 \end{tabular}}

\end{table*}

\begin{table*}[!t]\scriptsize
  \caption{Overview of approaches proposing Probabilistic and Markov chain-based Trust Management methods}
  \label{tab:probabilistic}
  \centering
\resizebox{\textwidth}{!}{\begin{tabular}{||c | c | c | c | c | c | c | c | c | c||} 
 \hline
 Paper & \makecell{Info \\ Gathering}   & Trust Update & Experiments & Centralized & \makecell{Trust \\ Formation} & \makecell{Threat \\ Model} & Simulator & Limitations & Strengths  \\ [0.5ex] 
 \hline\hline
 \cite{Boudagdigue} & Both  &  & \checkmark &  & Multi-Trust &\checkmark & & \makecell[l]{$\bullet$ Doesn't provide details for the experimental environment \\ $\bullet$ Deals with a set of trust-related attacks} &  \makecell[l]{$\bullet$ Improving an existing method} \\ 
 \hline
\cite{Wang} & Both &   & \checkmark& & Multi-Trust & & & \makecell[l]{$\bullet$ Do not specify the threat model}&\makecell[l]{$\bullet$ Dealing with the communication latency}  \\ 
 \hline
 \cite{Joshi} & Direct   & Time-Driven & \checkmark & & Multi-Trust & \checkmark & \checkmark &  \makecell[l]{$\bullet$ Deals with only 2 kind of attacks} &  \makecell[l]{$\bullet$ Takes into account the energy consumption}\\
 \hline
  \cite{WangMTES} & Direct    &   & \checkmark & & Multi-Trust &  & \checkmark & \makecell[l]{$\bullet$ Do not specify the threat model} &\makecell[l]{$\bullet$ Travel strategy for energy saving \\ $\bullet$ Mobility}\\
 \hline
    \cite{Fang} & Both  &  & \checkmark &  & & &\checkmark & \makecell[l]{$\bullet$ Do not specify the threat model \\ $\bullet$ Doesn't specify when a transaction is a success, fail or uncertain.} &  \makecell[l]{$\bullet$ Gives a solution for the cluster head selection}\\ 
 \hline
  \cite{FangFETMS} & Both  & Event-Driven & \checkmark &  & & \checkmark &\checkmark & \makecell[l]{$\bullet$ Only one attack is considered} &  \makecell[l]{$\bullet$ Cyber-security requirements for ICN} \\  [1ex]
 \hline
 \end{tabular}}
\end{table*}

\subsection{Probabilistic and Markov chain} 
There has been some research on estimating trust based on probability theory. A summary of the findings for this category is summarized in Table \ref{tab:probabilistic}.

We can observe from the Table that 33.3\% of the papers are using only Direct observations, while the rest of them (66.7\%) are also using recommendations for information gathering. Also, 16.7\% of the papers preferred a Time-Driven approach, 16.7\% an Event-driven one, and the rest of them did not refer to the Trust Update. All of the papers present Experiments, but 66.6\% of them refer to the Simulator used. Also, none of the papers is using a Centralized entity for trust evaluation.  Regarding Trust Formation, 66.6\% are using Multi-Trust, while the rest do not refer to the Trust Update. Finally, 50\% of them defined the Threat Model.

Boudagdigue et al. \cite{Boudagdigue} proposed a system where every node is monitored by its neighbors. Also, some groups of neighbors are formed to evaluate indirect trust. This paper proposes a distributed trust model based on a similar model proposed for vehicular networks. The authors used Markov Chains to model the trust change. A discrete-time chain with $M+1$ states was introduced. State $0$ corresponds to the lower trust level and $M$ to the uppermost. The probabilities were calculated based on the direct and indirect trust scores. The malicious nodes can perform BMA, BSA, selfish attacks, and honesty attacks. The trust parameters taken into account are honesty and cooperation. The experiments tested the proposed solution against different kinds of trust-related attacks. The results were presented as diagrams. One limitation of the current solution is that it only deals with a set of trust-related attacks. Also, the authors do not provide details about the experimental environment. On the other hand, the authors are improving an existing method proposed for VANETS in the IoT context.

Wang et al. \cite{Wang} proposed a trust model based on direct and indirect trust computation with trust prediction. The prediction method depends on the combination of exponential smoothing and a Markov chain. Exponential smoothing was employed to predict trust and a Markov chain was employed to fix any deviation. Thus, a prediction method was employed to predict the current trust level based on interaction history, behavior history, and some other factors like the device model. For the trust computation, both social and unsocial parameters were considered. The following experiments were performed: comparison of trust prediction with different exponential smoothing coefficients; comparison between first and second exponential smoothing; and experiments for different kinds of attacks. The results were presented as diagrams. The authors do not specify the threat model, which is a drawback. On the other hand, they are proposing a solution to the communication latency issue.

Joshi et al. \cite{Joshi} proposed a system that consists of several resource-constrained IoT nodes with a short radio range and a base station with a limitless source of energy as a central authority. This research work has presented a 2-state HMM with a Trusted state and a compromised state, together with essential and unessential output as observation states. The trustworthiness of the node is modeled by the 2-state HMM to predict the likelihood of the node's next state. The state transition probability matrix is defined by the energy consumed, the number of modified packets, and the number of forwarding packets. The malicious nodes can drop the packets or tamper with the data. Experiments were conducted in MATLAB to evaluate the network's trustworthiness with various percentages of compromised nodes and compare it with other methods. The results were presented in diagrams. The authors are taking into account only two kinds of attacks. The authors are using energy consumption as a key characteristic for calculating trust. This is interesting since increased activity might be malicious, but also energy of the nodes is also taken into account in a resource-constrained environment.

Wang et al. \cite{WangMTES} proposed a system consisting of Mobile edge nodes (MEN) and common sensors. The MEN are connected to a small number of sensors. In this paper, a mobile edge trust evaluation scheme is proposed. The evaluation of the trustworthiness of sensor nodes is achieved using a probabilistic graph model. The probabilistic graph model is used to represent the relationship between nodes. The interaction of node $i$ with node $j$ can be described as $P$ and $Q$. $P$ is a positive influence of node $i$ on node $j$ and $Q$ is a negative one. The information gathered for the formation of trust is the result of data collection and communication behavior. Also, a moving strategy method is proposed to decrease the travel distance MEN has to cover to evaluate every sensor. The experiments were conducted using MATLAB and NS-3 and were focused on the performance of the mechanism, the analysis of energy consumption, and the testing of the proposed moving algorithm. The results were presented in diagrams. The paper does not specify the threat model, which is a drawback. A strength, on the other hand, is that the authors propose a moving strategy for energy savings in a high-mobility environment.

Fang et al. \cite{Fang} proposed a system with a cluster-based architecture. The paper proposes a trust management scheme using Dirichlet Distribution. Both direct observations and third-party recommendations are considered to calculate the trust value of a node. This work is proposed to defend against internal attacks. Experiments conducted on MATLAB focus on the comparison with Beta distribution-based and Gaussian distribution-based performance experiments for OOA. The final results were presented in diagrams. One limitation of the solution is that the authors do not specify the threat model. Also, they do not specify when a transaction is successful or not. On the other hand, they propose a solution for cluster head selection.

Fang et al. \cite{FangFETMS} proposed a trust management technology that guards the system against OOA. Also, Beta distribution is used for the trust evaluation procedure. The authors also mentioned the cyber-security requirements for Information-Centric Networking (ICN). Experiments were conducted on MATLAB to observe the performance of the scheme and compare it with other techniques. One drawback of the method is that it only covers one attack. On the other hand, the authors are studying the cyber-security requirements for ICN systems.


\section{Challenges}\label{challenges}
Based on the above analysis, we are presenting some highlights of the vulnerabilities we observed.

\begin{itemize}
    \item \textit{Scalability:} We can observe from the above analysis that most of the works did not take into account the scalability factor when conducting the experiments. In extended IoT networks, where a huge number of sensors are connected and communicating with each other, a trust management system should be able to respond efficiently. Especially in dynamic networks, where the number of nodes is not fixed, a trust management scheme should be able to adapt to a growing amount of work.
    \item \textit{Privacy:} Privacy is a really important factor in every system. Especially for resource-constrained IoT devices. A trust management system may handle sensitive data to calculate and preserve the trust between two nodes. For example, the frequency of communication between the nodes. Also, blockchain technology can solve multiple problems, but a public blockchain that offers decentralization lacks privacy preservation. If some sensitive piece of information has to be exposed in a smart contract or stored on a blockchain, it is visible to all the participants.
    \item \textit{Context:} Context is really important for trust. Some individuals are to be trusted in specific contexts or circumstances. A malicious node may be trustworthy only in a specific context. Only five papers included in this survey take context into account to calculate trust. The highly heterogeneous IoT networks act differently in different contexts.
    \item \textit{Energy:} Energy is a really important factor for resource-constrained IoT devices. The research community hasn't extensively investigated the issue of designing a lightweight trust management system. We observed that a few of the works are conducting experiments to measure the energy consumption caused by the operation of trust management. 
    \item \textit{Attacks:} The trust management systems add some trust-related attacks to the threat model. These kinds of attacks should be tackled by the trust management system to be valuable for the IoT network. During the analysis, we saw that most of the works are mainly dealing with trust-related attacks. The attacks that gained the most attention from the research community are BMA, BSA, and SPA. The OOA and OSA have not been studied so extensively. However, a trust management scheme should be able to detect multiple attacks, not just trust-oriented ones. Especially, blockchain-based works that refer to the IoT nodes as lightweight nodes should take the EA into account. In future work, it will be nice to investigate trust management techniques that tackle trust-related attacks but also deal with a variety of other attacks.
    \item \textit{Mobility:} IoT can also be involved in high-mobility tasks (e.g. smart vehicles). Designing a trust-management system that can be adjusted to a high-mobility environment is important. There is not a lot of work involved in dealing with this issue.
    \item \textit{Cold-start problem:} The initialization of the trust values during bootstrapping is an issue that has to be addressed. This problem also occurs when a new node enters the network. This is an interesting and important issue that has to be addressed, especially in dynamic environments where nodes come and go constantly.
    \item \textit{Threat Model:} We have noticed that there are a lot of works that do not refer to the threat model. In our opinion, it is highly important to state the threat model and the attacks the proposed trust management technique is tackling. Different kinds of attacks are suitable for each system.
    \item \textit{Pre-trusted entities:} some approaches require some pre-trusted entities. This is an assumption that might not be applicable in real-life scenarios. In case of such an assumption, it should be clearly stated which are the pre-trusted entities and their role in the system. However, these approaches are not always applicable.
    \item \textit{Detailed analysis of technologies used:} It is important to give detailed information about the components used in the system. Especially in the case where other technologies are used, for example, blockchain and machine learning. The characteristics of the blockchain should be given in detail since a minor assumption on the reader's side can change the whole system.
    \item \textit{Defence Mechanisms:} If a malicious node is detected, there should be some defense mechanisms. Some works propose the exclusion of the node of the network. The research community should focus more on this subject to propose better alternatives.
    \item \textit{Filtering Recommendations:} Filtering recommendations are important in cases where indirect trust is enabled. Filtering recommendations can help prevent trust-related attacks.
    \item \textit{Edge and Cloud Architectures:} Nowadays, IoT is connected with Edge and Cloud computing. These architectures can be exploited to design trust management methods and take some computational and storage weight of the resource-constrained IoT devices.
    \item \textit{Data Trust:} Some works are taking into account the data trust. This concept can help design trust management techniques that take into account general attacks (not only trust-related). Data tampering can cause major issues with the functionality of an application (e.g. e-health). 
    \item \textit{Storage:} Another important issue is storage. In some cases, nodes have to hold huge amounts of trust-related information. There are some works dealing with this issue, proposing some efficient storage mechanisms, but the field needs further investigation.
    \item \textit{Network Traffic:} There are a number of techniques aimed at detecting malicious traffic in IoT networks (see~\cite{SHAFIQ2020433} and the references therein). We believe that those techniques could constitute the basis of new trust management methods that could use the traffic classification parameters as the basis of the trust and reputation metrics.  
    
\end{itemize}

\section{Conclusion}\label{conclusion}

Security and trust are critical in IoT systems since devices often run in potentially hostile environments. One way to assess the trustworthiness of a node is by using trust management techniques. Indeed, trust management in IoT has gained a lot of researchers' attention in the last decade resulting in a vast literature that is not easy to navigate. Motivated by this challenge, our work has addressed several research questions:

\begin{itemize}
\item{RQ1 - Which methods are currently used in the field?} Our work answers this question by providing a structured overview of a comprehensive set of the literature in the field. Our survey follows a well-disciplined approach to systematic literature reviews, which would allow any interested researcher to update and reproduce our main findings. The structure of our overview follows a classification based on categories and dimensions that have emerged from our preliminary analysis. Table~\ref{tab:multicolov} provides a bird's eye view of all the papers covered, which are summarized and discussed in Section~\ref{class}. The main outcome is not just a guide to existing works but it also highlights which classes of approaches are more predominant (e.g. social-based approaches using direct and indirect information-gathering techniques) and which ones are less explored (e.g. centralized approaches). 

\item{RQ2 - What is the threat model of those proposals?} We answer this question by considering a wide class of attacks identified in the literature on IoT security and how the proposed approaches covered in the survey relate to them (see Section~\ref{class}). Our main conclusion is that most proposals are not specific in the threat model being considered and that most of the works focus on data integrity and trust-related attacks. In general, the are lot of potential attacks are not covered extensively by the intended threat models. 

\item{RQ3 - What are the strengths and limitations of each proposal?} Our work answers this question in~\ref{class} by providing the strengths and limitations of each specific paper covered by the survey.  
It emerges from our study, for example, that many approaches proposed in the literature do not clearly specify the intended threat model, which makes it difficult to understand the assumptions under which the proposed approach will be effective. 

\item{RQ4 - What are the challenges and future directions?} We answer this question in Section~\ref{challenges}, where discuss a list of areas such as privacy, scalability, and bootstrapping, and the challenges related to trust management in IoT. One example of challenge and future direction regards the lack of use of precise threat models that emerged from our evaluation of the limitations of existing approaches. We believe that the discussion in Section~\ref{challenges} can help researchers find interesting areas in need of investigation and development.

\end{itemize}

Overall, we believe that our work can help readers decide on the methods and technologies that are more suitable for a particular IoT trust management mechanism and the challenges that should be considered when designing such a system. Our work can also support researchers in identifying future research avenues. An example would be the use of attack-defense trees to build solid trust management systems, containing all possible attacks and countermeasures.

\section*{Acknowledgment}

This work has been supported by Innovation Fund Denmark and the Digital Research Centre Denmark, through bridge project ``SIOT – Secure Internet of Things – Risk analysis in design and operation''.




%


\bibliographystyle{plain}
\bibliography{bibliography}







\end{document}